\documentclass[lettersize,journal]{IEEEtran}
\IEEEoverridecommandlockouts
\usepackage{amsmath}
\usepackage{cite}
\usepackage{amsmath,amssymb,amsfonts}
\usepackage[ruled,linesnumbered,lined]{algorithm2e}
\usepackage{graphicx}
\usepackage{textcomp}
\usepackage{xcolor}
\usepackage{bm}
\usepackage{booktabs}
\usepackage{amsthm,amssymb,amsfonts}
\usepackage{caption,subcaption}

\usepackage{float}

\SetKwInput{Input}{input}
\SetKwInput{Output}{output}
\newtheorem{definition}{Definition}
\usepackage{array}
\newcolumntype{L}[1]{>{\raggedright\let\newline\\\arraybackslash\hspace{0pt}}m{#1}}
\newcolumntype{C}[1]{>{\centering\let\newline\\\arraybackslash\hspace{0pt}}m{#1}}
\newcolumntype{R}[1]{>{\raggedleft\let\newline\\\arraybackslash\hspace{0pt}}m{#1}}

\newtheorem{proposition}{Proposition}

\def\BibTeX{{\rm B\kern-.05em{\sc i\kern-.025em b}\kern-.08em
		T\kern-.1667em\lower.7ex\hbox{E}\kern-.125emX}}

\begin{document}
	\title{Generalized Beyond-Diagonal RIS Architectures: Theory and Design via Structure-oriented Symmetric Unitary Projection}
	\author{Xiaohua Zhou, \textit{Graduate Student Member, IEEE}, Tianyu Fang, \textit{Graduate Student Member, IEEE}, Yijie Mao, \textit{Member, IEEE}, and Bruno Clerckx, \textit{Fellow, IEEE}
		\thanks{(\textit{Corresponding Author: Yijie Mao})
			\par Part of this work has been presented in 2025 IEEE International Conference on Communications (ICC) \cite{qstem}.
			\par This work has been supported by the National Nature Science Foundation of China under Grant 62571331. The work of Tianyu Fang was supported by the Research Council of Finland through 6G Flagship under Grant 346208 and through project DIRECTION under Grant 354901.
			\par X. Zhou and Y. Mao are with the School of Information Science and Technology, ShanghaiTech University, Shanghai 201210, China (e-mail:
			{zhouxh3, maoyj}@shanghaitech.edu.cn). 
			\par T. Fang is with Centre for Wireless Communications, University of Oulu, Finland (e-mail: tianyu.fang@oulu.fi).
			\par B. Clerckx is with the Department of Electrical and Electronic Engineering, Imperial College London, London SW7 2AZ, U.K. (e-mail: b.clerckx@imperial.ac.uk).
		}
	}
	
	\maketitle
	
	\thispagestyle{empty}
	\pagestyle{empty}
	\begin{abstract}
		Beyond-diagonal reconfigurable intelligent surface (BD-RIS), which enables advanced wave control through interconnection of RIS elements, are gaining growing recognition as a promising technology for 6G and beyond. However, the enhanced flexibility of BD-RIS in controlling the phase and amplitude of reflected signals comes at the cost of high circuit complexity. In this paper, we propose two novel BD-RIS architectures, namely, the stem-connected RIS and cluster-connected RIS, to explore trade-off between circuit complexity and performance. Specifically, the proposed stem-connected RIS is capable of achieving the same performance as fully-connected RIS while significantly reducing circuit complexity. The proposed cluster-connected RIS offers a unified framework that generalizes existing BD-RIS architectures—including single-connected, fully-connected, group-connected, tree-connected (arrowhead), and forest-connected (arrowhead) RISs—as special cases. This framework enables a much more flexible trade-offs between circuit complexity and system performance than existing ones. Based on the proposed BD-RIS architectures, we introduce a novel and generalized structure-oriented symmetric unitary projection method for designing the scattering matrix across all BD-RIS configurations. This method is effectively applied to solve the sum channel gain maximization problem and other utility-based optimization problems. 
		Numerical results demonstrate that the proposed stem-connected RIS is the simplest architecture that achieves optimal BD-RIS performance, while the cluster-connected RIS further enlarges the performance–complexity trade-off range.
		Furthermore, the proposed projection-based algorithms demonstrate high efficiency, achieving performance comparable to conventional alternative optimization-based approaches while significantly reducing computational complexity. This highlights their promising applicability in practical BD-RIS designs.
	\end{abstract}
	\vspace*{-0.3cm}
	\section{Introduction}\label{sec:intro}
	\subsection{Development of BD-RIS Architectures}
	Recently, beyond-diagonal reconfigurable intelligent surface (BD-RIS) has emerged as a promising technology for next-generation wireless communications \cite{RIS2,li2025tutorialbeyonddiagonalreconfigurableintelligent}. It generalizes the traditional RIS architecture introduced in \cite{Wu2020}, commonly known as single-connected or diagonal RIS. Specifically, traditional RIS employs a diagonal scattering matrix (also known as the passive beamforming matrix or phase shift matrix) where each unit independently modulates its incident signal without any energy coupling between elements. 
	 In comparison, BD-RIS introduces controlled inter-element coupling, allowing the scattering matrix to extend beyond the diagonal form. This introduces additional degrees of freedom in the scattering matrix design, thereby guaranteeing performance improvements over conventional RIS architectures. Prior studies have demonstrated that BD-RIS can significantly enhance wireless system performance, such as improvements in received power and sum-rate \cite{fang2022fully,fang2023low}, expanded spatial coverage \cite{li2023beyond}, and reduced resolution bit requirements \cite{li2025beyond}, etc.

	Based on the interconnection topology among RIS elements, BD-RIS architectures can be generally categorized into\footnote{In this work, we focus on the most widely adopted reciprocal and lossless BD-RIS architecture, characterized by symmetric and unitary scattering matrices. Other BD-RIS configurations exist, including: (i) non-reciprocal models with asymmetric scattering matrices~\cite{liu2024nonreciprocal,nonreciprocal2025}; (ii) lossy models incorporating dissipative impedance components in the reconfigurable network~\cite{Nerini2025lossy,peng2025lossydiagonalreconfigurableintelligent}; and (iii) reciprocal, lossless models with symmetric unitary scattering matrices~\cite{wu2019intelligent,fang2022fully,fang2023low,shen2021modeling,Matteo2024}. These alternative architectures worth further investigation in future research.
	} single-connected \cite{wu2019intelligent}, fully-connected \cite{shen2021modeling},  group-connected \cite{shen2021modeling}, tree-connected and forest-connected \cite{Matteo2024} structures, which are outlined as follows: 
	\begin{itemize}
		\item \textit{Single-connected RIS} \cite{wu2019intelligent} refers to the conventional diagonal RIS configuration, in which each RIS element is connected to its own individual load or impedance network. This architecture allows the elements to operate independently, resulting in minimal circuit complexity compared with other BD-RIS architectures. However, the lack of interconnection among elements restricts its ability to adjust signal propagation, thereby limiting overall flexibility and performance.
		\item \textit{Fully-connected RIS} \cite{shen2021modeling} refers to a BD-RIS architecture in which every RIS element is interconnected with all other elements. This configuration enables comprehensive inter-element interaction, providing maximum control over both the phase and amplitude of reflected signals. As a result, fully-connected BD-RIS achieves the highest performance gains over conventional diagonal RIS, but at the cost of significantly increased circuit complexity.
		\item \textit{Group-connected RIS} is introduced in \cite{shen2021modeling} to alleviate the high circuit complexity of fully-connected RIS. In this design, RIS elements are partitioned into distinct groups, with interconnections limited to elements within the same group. This leads to a block-diagonal scattering matrix, with reduced circuit complexity but also performance degradation compared to a fully-connected RIS.
		\item \textit{Tree-connected RIS}, proposed in \cite{Matteo2024} based on graph theory, employs a tree-structured graph to interconnect RIS elements. This design offers an improved trade-off between system performance and circuit complexity compared with the aforementioned single-connected, fully-connected and group-connected RISs. Although tree-connected RIS has been shown to be the simplest BD-RIS architecture capable of achieving the optimal performance in single-user multiple-input single-output (MISO) systems \cite{Matteo2024}, its performance also degrades in multi-user MISO settings compared with fully-connected RIS \cite{wu2024-ppADMM,wu2025band,li2025tutorialbeyonddiagonalreconfigurableintelligent}.
		\item \textit{Forest-connected RIS}, proposed in \cite{Matteo2024}, is designed based on a forest-structured graph to further reduce the complexity of tree-connected RIS while achieving a performance-complexity trade-off between single-connected RIS and tree-connected RIS. In this architecture, RIS elements are divided into groups, with each group forming a tree-connected architecture. However, forest-connected RIS still leads to noticeable performance degradation in multi-user MISO scenarios.
	\end{itemize}	

	To address the limitations of tree-connected RIS and forest-connected RIS, we propose a novel  \textit{stem-connected RIS}, aimed at identifying the simplest structure that can attain the optimal performance in BD-RIS aided multi-user MISO systems. Furthermore, we propose an additional BD-RIS architecture, \textit{cluster-connected RIS}, designed to further enhance the trade-off between system performance and circuit complexity.
	
	\subsection{ Development of Passive Beamforming Design Algorithms}
	With the advancement of BD-RIS technology, various algorithms have been proposed to design the passive beamforming matrix for enhancing system performance in BD-RIS assisted networks.  Initially, the quasi-Newton method \cite{shen2021modeling,fang2022fully} is introduced by vecterizing the passive beamforming matrix and reformulating the BD-RIS optimization problem into an unconstrained form. The resulting problem is solved using the fmincon toolbox. Another vectorization-based method is the penalty dual decomposition (PDD) approach \cite{yuyuan2023}, which obtains a closed-form expression after vecterization operation on the passive beamforming matrix.
	However, both of these methods are specifically tailored for fully- and group-connected RIS architectures. Subsequently, several other algorithms have been developed, including a manifold optimization algorithm that removes the symmetry constraint of BD-RIS \cite{hongyu2023,hongyu2024,hongyu2023jasc,yuyuan2023}, a two-stage method \cite{fang2023low}, and an alternating optimization (AO)-based projected successive linear approximation (PSLA) algorithm \cite{202501joint}. However, these methods all rely on projecting the solution onto the symmetric unitary space, and thus are limited to enforcing the BD-RIS constraint for fully- and group-connected architectures only. In general, the aforementioned algorithms are not applicable to the BD-RIS with different architectures such as tree- or forest-connected RISs, which might include structural constraint on the susceptance matrix. To address this, the partially proximal alternating direction method of multipliers (pp-ADMM) proposed in \cite{wu2024-ppADMM} offers a unified framework for passive beamforming across arbitrary BD-RIS architectures, however, it suffers from high average CPU time.  Overall, the development of a general and efficient passive beamforming matrix design algorithm for all BD-RIS  architectures remains an open and challenging problem.

	\subsection{Contributions}
	The main contributions of this paper are summarized as follows:
	\par First, we propose two novel BD-RIS architectures — \textit{stem-connected RIS} and \textit{cluster-connected RIS} — grounded in graph theory. The stem-connected RIS, proposed in the conference version of this work \cite{qstem}, is designed by dividing the RIS susceptance matrix into two subsets. One subset of RIS ports, defined as stems, are fully-connected to all other ports, while the remaining ports are only connected to the stems. We demonstrate that this approach can achieve the same performance as the fully connected RIS, with a much lower circuit complexity. As rigorously proved in \cite{wu2025band}, both this architecture and the band-connected RIS achieve optimal performance while maintaining theoretically minimal circuit complexity. Building on this novel architecture, we further propose the cluster-connected RIS, which combines the advantages of the stem-connected and group-connected architectures. The RIS elements are first partitioned into several groups. Within each group, a stem-based topology is employed.  Notably, the proposed cluster-connected RIS is a more general BD-RIS architecture that subsumes the existing BD-RIS configurations metioned in the introduction as special cases. This framework allows us to flexibly prioritize different performance aspects, such as reducing circuit complexity or enhancing system performance, based on specific design requirements. 
	
	\par Second, we propose a novel and generalized structure-oriented symmetric and unitary projection method that projects any matrix onto the feasible set defined by each BD-RIS architecture. Specifically, by utilizing Takagi decomposition and a novel transformation matrix to address the structural constraint, the original projection problem is reformulated as a system of linear equations, which is efficiently solved using the least squares method. To the best of our knowledge, this is the first work to introduce such a generalized projection framework that accommodates the diverse structural constraints of BD-RIS. Instead of relying on the conventional alternating optimization method for designing active and passive beamforming, we adopt a more efficient approach that maintains strong performance while significantly reducing the average CPU time.
	
	\par Third, we effectively apply the proposed projection method to solve the sum channel gain maximization problem and other utility-based optimization problems. Specifically, for the sum channel gain problem, we first relax the structural constraints of the BD-RIS and compute the optimal scattering matrix that achieves the theoretical upper bound. Then, we apply the proposed projection method to map this matrix onto the feasible set defined by the corresponding RIS architecture. A closed-form solution for the projection is derived, ensuring both efficiency and practicality. Building on this approach, we further propose low-complexity two-stage algorithms to address the utility-based joint active and passive beamforming design problem, which significantly reduce the overall computational complexity.
	
	\par Fourth, extensive simulations are conducted to evaluate the performance of the proposed BD-RIS architectures and projection algorithms. The results highlight the effectiveness of the proposed stem-connected RIS, particularly when the number of stems equals to or is greater than the maximum spatial multiplexing gain in a multi-user MISO channel. For such cases, the proposed projection-based algorithm enables the stem-connected RIS to achieve the same performance as the fully-connected RIS, while maintaining lower complexity. Furthermore, the proposed cluster-connected RIS is shown to provide an enlarged trade-off region between system performance and circuit complexity. Moreover, the proposed projection algorithm achieves comparable system performance to existing baselines while significantly reducing the average CPU time, demonstrating its efficiency.
	
	\subsection{Organizations and Notations}

	\par \textit{Organizations:} The remainder of this paper is structured as follows. Section~\ref{Sec:BD-RIS model} introduces and analyses the proposed BD-RIS architectures. Section~\ref{Sec:sysmod} introduces the BD-RIS assisted multi-user communication model. Section~\ref{Sec:projection} presents the proposed structure-oriented symmetric unitary projection method. Simulation results are provided in Section~\ref{Sec:simulation}, followed by the conclusion in Section~\ref{Sec:conclusion}.
	
	\par \textit{Notations:} Bold upper and lower case letters denote matrices and column vectors, respectively, $ (\cdot)^T $, $ (\cdot)^H $, $ |\cdot| $, $ \| \cdot \| $, $ \mathbb{E}(\cdot) $, $ \operatorname{tr}(\cdot) $ and $ [\mathbf{A}]_{i,j} $ represent the transpose, Hermitian, absolute value, Euclidean norm, expectation, trace operators and the $ (i,j) $-th element of matrix $ \mathbf{A} $, respectively.

	\section{Proposed BD-RIS Modeling and Analysis}\label{Sec:BD-RIS model}
	
	In this section, we begin with an overview of existing BD-RIS architectures, then introduce two novel BD-RIS structures, followed by an analysis and comparison of their circuit complexity.

	\subsection{Existing BD-RIS Architectures}\label{sec:existing BD-RIS}
	The BD-RIS is modeled based on microwave network theory, where $N$ passive elements are interpreted as antennas interconnected via an $N$-port reconfigurable impedance network with tunable passive components \cite{shen2021modeling, pozar2021microwave}\footnote{This paper focuses exclusively on BD-RIS operating in reflection mode, where signals are received and reflected on the same side, for simplicity. While the proposed approaches and algorithms are designed for this configuration, they can be extended to advanced architectures such as STAR-RIS \cite{STARRISxu} and multi-sector BD-RIS\cite{hongyu2023jasc}, which are left for future work.}.  The behavior of this reconfigurable impedance network is characterized by its admittance\footnote{The use of admittance parameters over impedance or scattering parameters has some benefits to describe BD-RIS architectures as discussed in \cite{li2025tutorialbeyonddiagonalreconfigurableintelligent}.} parameters (also known as Y-parameters \cite{pozar2021microwave}), which is a symmetric purely imaginary admittance matrix $\mathbf{Y} = j\mathbf{B}$, where $\mathbf{B} \in \mathbb{R}^{N \times N}$ is the susceptance matrix satisfying
	\begin{equation}\label{eq:B-equ}
		\mathbf{B} = \mathbf{B}^T.
	\end{equation}
	This configuration ensures the BD-RIS is both reciprocal and lossless. The corresponding passive beamforming matrix (also known as the scattering matrix) $\bm{\Theta}$ of the BD-RIS can be directly derived from the susceptance matrix $\mathbf{B}$ as:
	\begin{equation} \label{eq:theta}
		\bm{\Theta} = \left( \mathbf{I}_N + j Z_0 \mathbf{B} \right)^{-1} \left( \mathbf{I}_N - j Z_0 \mathbf{B} \right),
	\end{equation}
	where $Z_0$ is the reference impedance.
	\par 
	With equations \eqref{eq:B-equ} and \eqref{eq:theta}, the resulting scattering matrix $\bm \Theta$ is guaranteed to be symmetric and unitary, satisfying the fundamental constraints of reciprocity and power conservation, i.e.,     \begin{subequations}\label{scatter}
		\begin{align}
			&\mathbf \Theta\mathbf \Theta^H=\mathbf I_N,\\
			&\mathbf \Theta=\mathbf \Theta^T.
		\end{align}
	\end{subequations}
	Note that \eqref{scatter} is equivalent to \eqref{eq:B-equ} and \eqref{eq:theta}, and therefore, most existing works on BD-RIS directly optimize the scattering matrix to approach the performance upper bound achievable with a fully connected architecture. However, when additional structural constraints are imposed on the susceptance matrix $\mathbf B$, any closed-form expression for the corresponding scattering matrix must explicitly involve $\mathbf B$. As a result, the design focus shifts from optimizing the scattering matrix to directly optimizing the susceptance matrix $\mathbf B$ which fully characterizes the physical interconnections between BD-RIS elements.
	%

	For different BD-RIS architectures, the varying topologies of the underlying impedance network may impose distinct additional structural constraints on $ \mathbf{B} $, as summarized below:
	\begin{itemize}
		\item \textbf{Fully-connected RIS:} All ports are interconnected via tunable impedance components, resulting in a susceptance matrix that satisfies
		\begin{equation}\label{eq:B-fully}
			\begin{split}\mathcal{B}_{\text{Fu}}= \left\{\mathbf{B}|
				\mathbf{B} = \mathbf{B}^T\right\}.
			\end{split}
		\end{equation}
		With \eqref{eq:theta} and \eqref{eq:B-fully},  the corresponding passive beamforming matrix should satisfy 
		\begin{equation}
			\left\{\bm \Theta | \bm \Theta =\bm \Theta^T, \; \bm \Theta \bm \Theta^H=\mathbf{I}\right\}.
		\end{equation}
		\item \textbf{Single-connected RIS:} This refers to the conventional diagonal RIS, in which the susceptance matrix B becomes diagnal as:
		\begin{equation}\label{eq:B-single}
			\mathcal{B}_{\text{Si}} = \left\{\mathbf{B}|\mathrm{diag}([\mathbf{B}]_{1,1}, \cdots, [\mathbf{B}]_{N,N})\right\}.
		\end{equation}
		Using \eqref{eq:theta} and \eqref{eq:B-single}, it follows that the correcponding passive beamforming matrix should satisfy:
		\begin{equation}
			\left\{ \bm \Theta | \bm \Theta = \operatorname{diag}\left(e^{j \theta_1},e^{j \theta_2},\cdots,e^{j \theta_N}\right) \right\},
		\end{equation}
		where $\theta_n \in [0,2\pi)$ denotes the phase shift angle, as considered in many existing works \cite{wu2019intelligent,qingqing2021}.
		\item \textbf{Group-connected RIS:} This BD-RIS architecture uniformly divides the RIS elements into $ G $ groups, where elements within each group are fully connected. As a result, the susceptance matrix $ \mathbf{B} $ becomes block-diagonal and symmetric, which is given as:
		\begin{equation}\label{eq:B-group}
			\begin{split}
				& \mathcal{B}_{\text{Gr}(G)}\!=\!\left\{\mathbf{B}|\mathrm{blkdiag}(\mathbf B_{1},\cdots,\mathbf B_{G}), \mathbf{B}_g\!=\!\mathbf{B}_g^T,  g\!\in\mathcal C\right\},
			\end{split}
		\end{equation}
		where $\mathcal C\triangleq\{1, \cdots, G\}$ and $ \mathbf{B}_g$ is the symmetric susceptance matrices in group $g$. With \eqref{eq:theta} and \eqref{eq:B-group},  the corresponding passive beamforming matrix should satisfy 
		\begin{equation}\small
			\left\{ \bm \Theta | \bm \Theta = \operatorname{blkdiag} (\bm \Theta_1, \cdots, \bm \Theta_G), \bm \Theta_g = \bm \Theta_g^T, \bm \Theta_g\bm \Theta_g^H=\mathbf{I}_{N_g} \right\},
		\end{equation}
		where $N_g$ denotes the number of elements within each group, and $\bm \Theta_g, g \in \mathcal C$ are complex symmetric unitary matrices.
		When $ G=1 $, this structure reduces to fully-connected RIS. When $ G=N $, this structure reduces to single-connected RIS.
		
		\item \textbf{Tree-connected RIS (arrowhead):} This architecture is directly defined through the susceptance matrix $ \mathbf{B} $, enabling the circuit topology to form a tree structure in the sense of graph theory. The corresponding $ \mathbf{B} $ should satisfy \cite{Matteo2024}: 
		\begin{equation}\label{eq:B-tree}
			\begin{split}  
				 \mathcal{B}_{\text{Tr}} = &\left\{ \mathbf{B} \,\middle|\, 
				[\mathbf{B}]_{n,m} = 0 \text{ for } n \neq m,\;n > 1,\; m > 1, \right. \\
				&
				 \left. \mathbf{B}=\mathbf{B}^T
				\right\}.
			\end{split}
		\end{equation}
		\item \textbf{Forest-connected RIS (arrowhead):} This BD-RIS architecture combines the features of arrowhead tree-connected and group-connected RIS by dividing the elements into $ G $ groups, where the elements within each group are connected following a tree topology.  Consequently, the susceptance matrix $ \mathbf{B} $ satisfy \cite{Matteo2024}:
		\begin{equation}\label{eq:B-forest}
			\begin{split}
				 \mathcal{B}_{\text{Fo}(G)} =&\left\{\mathbf{B}| \mathrm{blkdiag}(\mathbf{B}_1, \cdots, \mathbf{B}_G),\right.\\ 
				 &\left. \mathbf{B}_g = \mathbf{B}_g^T, \mathbf{B}_g 
				 \in \mathcal{B}_{\text{Tr}}\right\},
			\end{split}
		\end{equation}
		where $ \mathcal{B}_{\text{Tr}} $ is defined in \eqref{eq:B-tree}.
	\end{itemize}
	 As highlighted in Section \ref{sec:intro}, although tree-connected and forest-connected RIS architectures demonstrate improved trade-offs between system performance and circuit complexity in single-user single-input single-output (SISO) scenario \cite{nerini2023pareto}, there remains a lack of a structure that achieves the optimal balance in multi-user MISO. Moreover, there is a need for a general and versatile intelligent surface framework that encompasses the aforementioned RIS architectures as special cases. In the following subsections, we propose two novel BD-RIS architectures to address this research gap.
	
	\subsection{Proposed Stem-connected RIS}\label{sec:stem-con}
	We begin with a novel BD-RIS architecture termed the stem-connected RIS, which integrates key characteristics of single-connected, fully-connected, and tree-connected RIS structures. Utilizing graph-theoretical tools, we model the general circuit topology of the proposed stem-connected RIS in Definition \ref{def:stem}.
	\begin{definition}[\textbf{Stem-connected RIS}]\label{def:stem}
		An undirected graph $\mathcal{G} = (\mathcal{V}, \mathcal{E})$ is employed to describe the inter-port topology for stem-connected RIS, where $\mathcal{V} = \{1, 2, \cdots, N\}$ is the vertex set corresponding to the RIS ports, and $\mathcal{E}$ is the edge set and each edge $(n,m)\in \mathcal{E}$ represents a connection between RIS ports $n$ and $m$ through a tunable admittance element.
		
		Define $Q (Q < N)$ as the number of stems, the RIS ports in $\mathcal{V}$ are divided into two subsets:
		\begin{equation*}
			\mathcal{V}_Q = \{1, \cdots, Q\}, \;
			\mathcal{V}_{\Bar{Q}} = \{Q+1, \cdots, N\}.
		\end{equation*}
		Stem-connected RIS enables ports in $\mathcal{V}_Q$ to connect with all other ports in $\mathcal{V}$, while ports in $\mathcal{V}_{\Bar{Q}}$ only connect to those in $\mathcal{V}_Q$. This interconnection rule defines the edge set of the stem-connected RIS as:
		\begin{equation}\label{eq:edge-Q}
			\begin{aligned}
				\mathcal{E} = \{(n,m) \,|\, &n \in \mathcal{V}_Q, m \in \mathcal{V}, \text{ or } n \in \mathcal{V}_{\Bar{Q}}, m \in \mathcal{V}_Q; \\
				&[\mathbf{B}]_{n,m} \neq 0, \, n \neq m \}.
			\end{aligned}
		\end{equation}
		Here, $ [\mathbf{B]}_{n,m} \neq 0 $ indicates that a tunable admittance exists between the $n$-th and the $m$-th ports.
	\end{definition}
	
	\begin{figure}[t!]
		\centering
		\begin{subfigure}{1\linewidth}
			\centering 
			\includegraphics[width=0.7\linewidth]{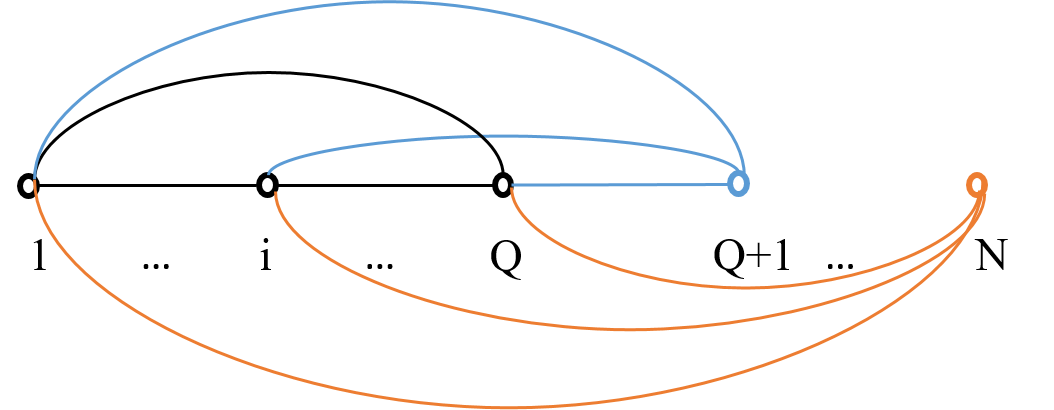}
			\caption{Graph representation for stem-connected RIS.}
			\label{fig:graph}
		\end{subfigure}
		\vspace{1mm}
		\begin{subfigure}{1\linewidth}
			\centering
			\includegraphics[width=0.7\linewidth]{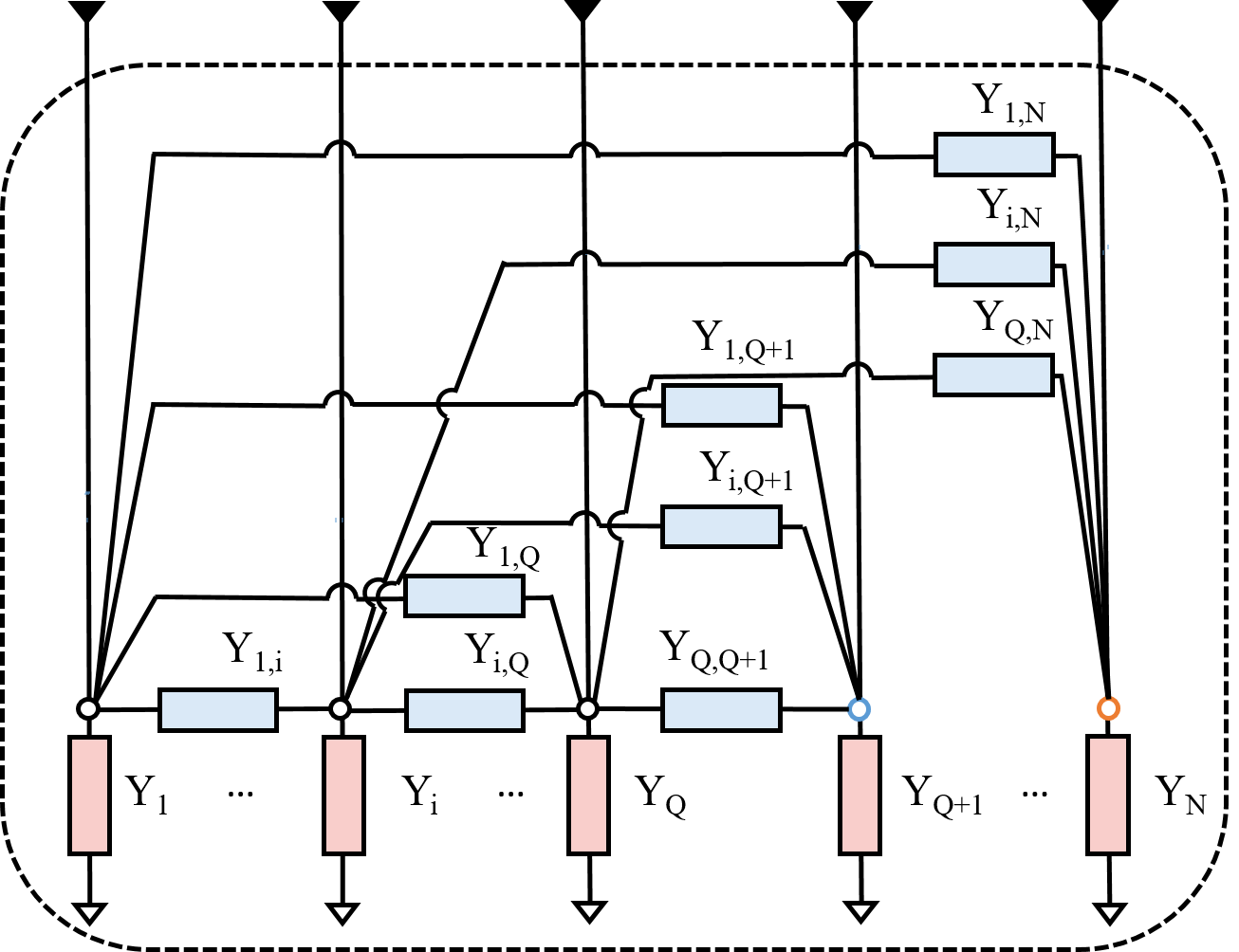}
			\caption{The architecture of the proposed $N$-port stem-connected RIS.}
			\label{fig:port}
		\end{subfigure}
		\label{fig:Q-stem-architecture}
		\caption{The circuit architecture of stem-connected RIS and its graph representation.}
	\end{figure}

	The graph representation for the stem-connected RIS is illustrated in Fig. \ref{fig:graph}. According to Definition \ref{def:stem}, the susceptance matrix $ \mathbf{B} $ of stem-connected RIS should satisfy the following sparsity constraint:
	\begin{equation}\label{eq:B-arrow}\small
		\mathcal{B}_{\text{St}(Q)} = \left\{ \mathbf{B} \,|\, [\mathbf{B}]_{n,m} = 0 \text{ for } n > Q, m > Q, n \neq m, \mathbf{B}=\mathbf{B}^T \right\}.
	\end{equation}
	
	\par Given that $\mathbf{Y}=j\mathbf{B}$, the admittance matrix $ \mathbf{Y} $ of the stem-connected RIS should follow 
	\begin{equation}\label{eq:Y-arrow}
		\left\{ [\mathbf{Y}]_{n,m} = 0 \text{ for } n > Q, m > Q, n \neq m, \mathbf{Y}=\mathbf{Y}^T\right\},
	\end{equation}
	where $[\mathbf{Y}]_{n,m}$ denotes the tunable admittance connecting the $n$-th port to the $m$-th port. Therefore,  the overall circuit architecture of our proposed stem-connected RIS is obtained as illustrated in Fig. \ref{fig:port}, where the first $Q$ ports are connected to all other ports while the remaining ports are only connected to the first $Q$ ports. 
	
	
	\begin{figure}
		\centering
		\includegraphics[width=0.6\linewidth]{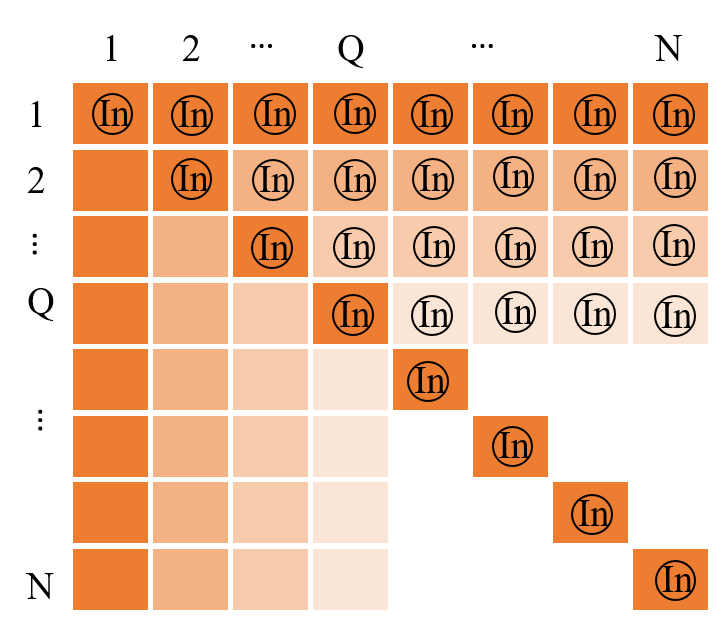}
		\caption{The shape of a feasible
			susceptance matrix $ \mathbf{B} $ for stem-connected RIS. It satisfies $\mathbf{B} \in \mathcal{B}_{\text{St}(Q)}$ and $\mathbf{B}=\mathbf{B}^T$. The non-zero elements in $\mathbf{B}$ are represented in different shades of orange. The upper triangular part of the matrix $\mathbf{B}$ is labeled as ``\textcircled{In}''.
		}
		\label{fig:arrow}
		\vspace{-4mm}
	\end{figure}

	Fig. \ref{fig:arrow} further illustrates the shape of a feasible susceptance matrice $ \mathbf{B} $ satisfying constraint \eqref{eq:B-arrow}.  Notably, by flexibility tuning the value of $Q$, the proposed stem-connected RIS framework enables a tradeoff between circuit complexity and performance, and also generalizes existing single-connected, tree-connected, and fully-connected RIS as special cases. Specifically:
	\begin{itemize}
		\item  When $Q=0$, the stem-connected RIS reduces to a single connected RIS, satisfying \eqref{eq:B-single}.
		\item When $Q=1$, it becomes a tree-connected RIS, satisfying \eqref{eq:B-tree}.
		\item $Q=N-1$: When $Q=N-1$, it corresponds to a fully-connected RIS, satisfying \eqref{eq:B-fully}.
	\end{itemize}

	Our proposed stem-connected RIS cannot generalize group-connected RIS or forest-connected RIS architectures. The primary reason is that a group-connected RIS partitions the RIS elements into multiple groups, resulting in a block-diagonal susceptance matrix. This structure inherently leads to a block-diagonal scattering matrix, which simplifies the circuit design but limits the system's ability to fully exploit inter-element coupling. While this approach reduces circuit complexity, it comes at the cost of degraded performance compared to a fully-connected RIS if the group size exceeds one. In contrast, the proposed stem-connected RIS employs a sparse susceptance matrix but still achieves a full scattering matrix. By carefully controlling the number and placement of non-zero entries in the susceptance matrix, it can achieve the performance of a fully-connected RIS while significantly reducing circuit complexity.
	This will be further illustrated in the simulation section.

	
	\subsection{Cluster-connected RIS}
	Inspired by the structural extension from tree-connected RIS to forest-connected RIS, we further propose a cluster-connected RIS in this subsection, which combines the key features of stem-connected and group-connected architectures. Specifically, the RIS elements are divided into $ G $ groups, each containing $N_G$ elements, such that $GN_G=N$. Within each group, the elements are connected according to a stem topology. This configuration enables a more generalized BD-RIS architecture that encompasses all five existing RIS configurations discussed in Section \ref{sec:existing BD-RIS} as special cases.
	
	\begin{definition}[\textbf{Cluster-connected RIS}]\label{def:cluster}
		By dividing the RIS elements into $ G $ groups, each containing $N_G$ elements such that $GN_G=N$, the circuit topology of a cluster-connected RIS is represented by an undirectional graph $\mathcal{G}=(\mathcal{V},\mathcal{E})$, where the vertex and edge sets are defined as
		\begin{equation}
			\mathcal{V}= \cup_{g=1}^{G} \mathcal{V}_g, \mathcal{E}=\cup_{g=1}^{G} \mathcal{E}_g.
		\end{equation}
		Here, $\mathcal{V}_g$ and $\mathcal{E}_g$ respectively denote the vertex and edge sets of the $g$-th group. Specifically, the vertex set of the $g$-th group is given by
		\begin{equation*}
			\begin{split}
				\mathcal{V}_g = \left\{(g-1)N_G+1, (g-1)N_G+2, \dots, (g-1)N_G+N_G \right\}, \\
				\forall g \in \{1,2, \dots, G\}.
			\end{split}
		\end{equation*}
		
		Furthermore, by defining $Q_G (Q_G < N_G)$ as the number of stems in each group, the vertex set $V_g$ of group $g$ is further partitioned into two subsets:
		\begin{equation*}\small
			\begin{split}
				\mathcal{V}_g^{Q_G} = \left\{(g-1)N_g+1, (g-1)N_g+2, \dots, (g-1)N_g+Q_G \right\}, \\
				\forall g \in \{1,2, \dots, G\},
			\end{split}
		\end{equation*}
		\begin{equation*}
			\begin{split}
				\mathcal{V}_g^{\Bar{Q}_G} = \mathcal{V}_g \backslash \mathcal{V}_g^{Q_G}.
			\end{split}
		\end{equation*}
		The cluster-connected RIS allows the vertices within each group to be connected following a stem topology. Specifically, the vertices in
		$ \mathcal{V}_g^{Q_G} $ are connected to all other vertices in
		$ \mathcal{V}_g $, while the remaining vertices in $ V_g^{\bar{Q}_G} $ are connected only to those in $ \mathcal{V}_g^{Q_G} $. Consequently, the edge set for each group is given by:
		\begin{equation}\label{eq:edge-QG}
			\begin{split}
				\mathcal{E}_g = \left\{(n,m) \mid n \in \mathcal{V}_g^{Q_G}, m \in \mathcal{V} \text{ or } n \in \mathcal{V}_g^{\bar{Q}_G}, m \in \mathcal{V}_g^{Q_G}, \right. \\
				\left. [\mathbf{B}]_{n,m} \neq 0, n \neq m \right\}.
			\end{split}
		\end{equation}
	\end{definition}
	\begin{figure}
		\centering
		\includegraphics[width=0.6\linewidth]{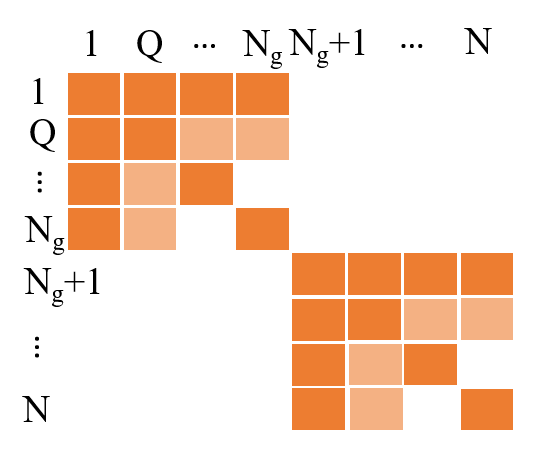}
		\caption{The shape of a feasible
			susceptance matrix $ \mathbf{B} $ for cluster-connected RIS. Each $\mathbf{B}_g \in \{\mathbf{B}_1, \cdots, \mathbf{B}_G\}$ satisfies $\mathbf{B}_g \in \mathcal{B}_{Q_G,g}$ and $\mathbf{B}_g=\mathbf{B}_g^T$. 
		}
		\label{fig:cluster}
		\vspace{-4mm}
	\end{figure}
	Following Definition \ref{def:cluster}, the susceptance matrix $\mathbf{B}$ for cluster-connected RIS should be a block-diagonal matrix, i.e., 
	\begin{equation}\label{eq:B-cluster}
		 \mathcal{B}_{\text{Cl}(Q_{G},G)}=\left\{\mathbf{B}|\mathrm{blkdiag}\{\mathbf{B}_1, \mathbf{B}_2,...,\mathbf{B}_G\}, \mathbf{B}_g \in \mathcal{B}_{Q_G,g} \right\},
	\end{equation} 
	where $\mathbf{B}_g$ denotes the $g$-th diagonal block. The feasible set of $ \mathbf{B}_g $, denoted by $\mathcal{B}_{Q_G,g}$, is characterized as
	\begin{equation}\label{eq:B-cluster-g}
		\begin{split}
			\mathcal{B}_{Q_G,g} = \left\{\mathbf{B}_g \,\middle|\, [\mathbf{B}_g]_{n,m} = 0,\ \forall n \neq m, \right.\\  n > Q_G,\ m > Q_G, 
			\left.\mathbf{B}_g=\mathbf{B}_g^T\right\}.
		\end{split}
	\end{equation}
	Fig.~\ref{fig:cluster} illustrates a toy example of a feasible susceptance matrice for cluster-connected RIS when $N=8$, $G=2$ and $Q_G=2$. By flexibility tuning $G$ and $Q_G$, the proposed cluster-connected RIS framework offers an enhanced trade-off between circuit complexity and performance, while also unifying all previously discussed RIS architectures as special cases. Specifically:
	\begin{itemize}
		\item When $G=1$, the architecture reduces to a stem-connected RIS, satisfying \eqref{eq:B-arrow}.
		\item When $Q_G=N_G-1$, it corresponds to a group-connected RIS, satisfying \eqref{eq:B-group}.
		\item When $Q_G=1$, it becomes a forest-connected RIS, satisfying \eqref{eq:B-forest}.
		\item  When $G=1$ and $Q_G=0$, the cluster-connected RIS reduces to a single-connected RIS, satisfying \eqref{eq:B-single}.
		\item When $G=1$ and $Q_G=N-1$, it corresponds to a fully-connected RIS, satisfying \eqref{eq:B-fully}.
		\item When $G=1$ and $Q_G=1$, it becomes a tree-connected RIS, satisfying \eqref{eq:B-tree}.
	\end{itemize}
	This hierarchical structure highlights the versatility of the cluster-connected architecture in adapting to various design requirements. It is worth noting that unlike the dynamically connected RIS proposed in~\cite{li2023dynamic,li2025tutorialbeyonddiagonalreconfigurableintelligent}, which adjusts impedance values and circuit interconnections dynamically, our proposed cluster-connected RIS adopts a fixed yet general architecture. It thereby provides a more practical solution to generalize the aforementioned BD-RIS architectures without incurring additional hardware costs and tuning delay.
	
%

	\addtolength{\topmargin}{0.03in}

	\subsection{Circuit Complexity Analysis}
\begin{figure}
		\centering
		\includegraphics[width=0.8\linewidth]{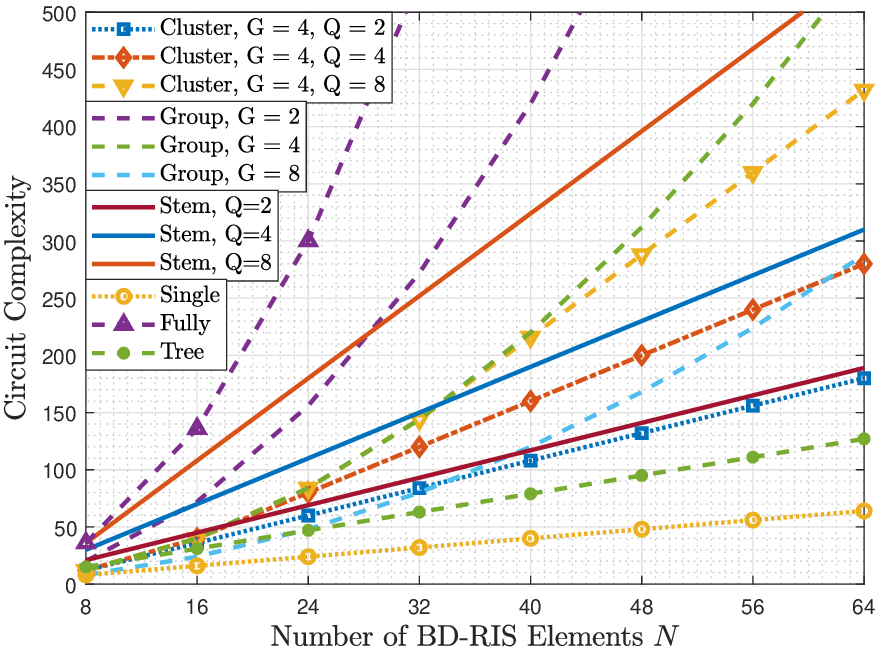}
		\caption{Circuit topology complexity versus the number of BD-RIS elements.}
		\label{fig:circuit}
		\vspace{-4.5mm}
	\end{figure}

	The circuit complexity is determined by the number of independent tunable admittance components, which is equivalent to the number of non-zero elements in the symmetric susceptance matrix $\mathbf{B}$. 
	For the stem-connected RIS, $N$ admittance components are required to connect each RIS port to the ground. For inter-port connections as illustrated in Fig.~\ref{fig:port}, the 1st port connects to $N - 1$ others, the 2nd one connects to $N - 2$, and so on, up to the $Q$-th port which connects to $N - Q$ others. This results in $QN - \frac{Q(Q + 1)}{2}$ admittance components interconnecting the ports. Hence, the total number of admittance components in the stem-connected RIS is
	$ QN + N - \frac{Q(Q + 1)}{2} $.
	For the cluster-connected RIS with $G$ groups, the stem-connected architecture is employed within each group. As a result, each group contains $Q_G N_G - \frac{Q_G(Q_G + 1)}{2}$ admittance components interconnecting the ports. Since $N_G G = N$, the total circuit complexity of the cluster-connected RIS is $ Q_G N + N - G \frac{Q_G(Q_G + 1)}{2} $.
	 Fig.~\ref{fig:circuit} further illustrates circuit complexity versus the number of RIS elements across different RIS architectures. We observe that the circuit complexities of the fully-connected and group-connected RISs grow quadratically with respect to $N$, while the circuit complexities of the single-connected, tree-connected, forest-connected, stem-connected, and cluster-connected RISs grow linearly with $N$ for fixed stem numbers $Q$ and $Q_G$. Furthermore, the stem-connected RIS demonstrates significantly lower circuit complexity compared to the fully-connected RIS, while the cluster-connected RIS achieves even lower complexity than the stem-connected RIS. This highlights the superior scalability of the two architectures for large-scale RIS deployments. 


	\section{BD-RIS Assisted Multi-user Communication}\label{Sec:sysmod}

	To assess the effectiveness of the proposed stem-connected and cluster-connected RIS architectures, this section considers a BD-RIS-assisted multi-user multiple-input single-output (MU-MISO) downlink communication system and formulates the associated beamforming design problems, which will be addressed in the subsequent sections.
	\vspace{-4.5mm}
	\subsection{System Model}\label{sec:sys-mod}
	We consider a BD-RIS-assisted multi-user downlink communication system, where a base station (BS) equipped with $L$ transmit antennas simultaneously serves $K$ single-antenna users via a BD-RIS comprising $N$ passive reflecting elements. Let $\mathcal{N} = \{1, \cdots, N\}$ and $\mathcal{K} = \{1, \cdots, K\}$ denote the index sets of RIS elements and users, respectively. The BD-RIS is characterized by the scattering matrix $\bm \Theta \in \mathbb{C}^{N \times N}$. For different BD-RIS architectures, $ \bm \Theta $ must satisfy different constraints. For instance, in the proposed stem-connected RIS, $ \bm \Theta $ should comply with constraint \eqref{eq:B-arrow}. In contrast, for the proposed cluster-connected RIS, $ \bm \Theta $ must satisfy \eqref{eq:B-cluster}. Assuming the direct BS-user links are blocked, all users are served exclusively through the BS-to-BD-RIS and BD-RIS-to-user channels. Let $\mathbf{E} \in \mathbb{C}^{N \times L}$ denote the channel matrix from the BS to the BD-RIS, and $\mathbf{h}_k \in \mathbb{C}^{N \times 1}$ denote the channel vector from the BD-RIS to user $k$. The resulting cascaded (effective) channel from the BS to user $k$ is given by $\mathbf{f}_k^H = \mathbf{h}_k^H \bm{\Theta} \mathbf{E} \in \mathbb{C}^{1\times L}$. By defining $\mathbf{H} = [\mathbf{h}_1, \cdots, \mathbf{h}_K] \in \mathbb{C}^{N \times K}$, the aggregated effective channel from the BS to all users becomes $\mathbf{H}^H \bm{\Theta} \mathbf{E}$.

	
	To serve all users simultaneously, the BS employs linear beamforming to shape and direct the transmit signal towards the users.
	Let $s_k \in \mathbb{C}$ denote the complex-valued symbol intended for user $k$, with $\mathbb{E}[|s_k|^2] = 1$, and $\mathbf{w}_k \in \mathbb{C}^{L \times 1}$ be the corresponding active beamforming vector. The transmit signal at the BS is $ \mathbf{x} = \sum_{k=1}^K \mathbf{w}_k s_k $.
	The signal received at user $k$ is then given by $y_k = \mathbf{f}_k^H \sum_{i=1}^K \mathbf{w}_i s_i + n_k$,
	where $n_k \sim \mathcal{CN}(0, \sigma_k^2)$ denotes the additive white Gaussian noise.
	Define $ \mathbf{y} = [y_1, \ldots, y_K]^T $, $\mathbf{s} = [s_1, \ldots, s_K]^T \sim \mathcal{CN}(\mathbf{0}, \mathbf{I}_K)$, $\mathbf{W} = [\mathbf{w}_1, \ldots, \mathbf{w}_K] \in \mathbb{C}^{L \times K}$, and $\mathbf{n} = [n_1, \ldots, n_K]^T \in \mathbb{C}^{K \times 1}$. The collective signal received at all users is:
	\begin{equation}
		\mathbf{y} = \mathbf{H}^H \bm{\Theta} \mathbf{E} \mathbf{W} \mathbf{s} + \mathbf{n}.
	\end{equation}
	
	\vspace{-4.5mm}
	\subsection{Problem Formulation}\label{sec:channel-gain}
	In this work, we focus on designing the scattering matrix $ \bm \Theta $ by maximizing the effective channel gain between the BS and all users. The proposed algorithm is subsequently extended to address a more general joint optimization problem involving both active and passive beamforming.

	The channel gain optimization problem, applicable to any type of BD-RIS architecture, can be generally formulated as:
	\begin{subequations}\label{eq:channelgain-formula}
		\begin{align}
			\max_{\bm \Theta} \quad &\|\mathbf H^H\mathbf \Theta\mathbf E\|_F^2 \label{subeq:obj}\\
			\operatorname{s.t.} \quad &\bm \Theta=\left(\mathbf{I}+jZ_0\mathbf{B}\right)^{-1}\left(\mathbf{I}-jZ_0\mathbf{B}\right), \label{eq:theta-con}\\
			&\mathbf{B} \in \mathcal{B}_x.\label{eq:B-stru}
		\end{align}
	\end{subequations}
	where $x\in\{\text{Si}, \text{Tr}, \text{Fu}, {\text{Gr}(G)}, {\text{Fo}(G)}, {\text{St}(Q)},  {\text{Cl}(Q_G,G)}\}$, depending on the specific BD-RIS architecture under consideration. The problem is non-convex due to the nonlinear transformation of the susceptance matrix $ \mathbf{B} $ in the scattering matrix $\bm \Theta $, which involves a matrix inversion that is inherently non-convex. Moreover, the quadratic objective function $\|\mathbf H^H\mathbf \Theta\mathbf E\|_F^2$ further complicates the optimization, as $ \bm \Theta $ is already a nonlinear function of $ \mathbf{B} $. Most importantly, the structural constraint in \eqref{eq:B-stru} introduces combinotoral nature to the problem, making the feasible set highly non-convex.
	
	\par Prior works have studied this channel gain maximization problem for fully-connected RIS~\cite{takagi2023,fang2023low}, tree-connected RIS~\cite{Matteo2024}, and group-connected RIS~\cite{nerini2023closed}. However, these algorithms are not directly applicable to our proposed stem-connected and cluster-connected RIS due to their unique structural constraints on the scattering matrix. Moreover, the algorithms proposed in~\cite{takagi2023,nerini2023closed,Matteo2024} are tailored for single-user scenarios.
	In the following, we first specify a theoretical upper-bound solution to problem \eqref{eq:channelgain-formula} by relaxing constraint \eqref{eq:B-stru}, and then bridge the research gap by proposing a general projection method compatible with the structural constraints of various BD-RIS architectures.
	
	\subsection{Theoretical Upper Bound Analysis for Problem \eqref{eq:channelgain-formula}}\label{sec:Upperbound}
	By removing constraint \eqref{eq:B-stru} and applying  singular value decomposition (SVD) to $ \mathbf{H}^H $ and $ \mathbf{E} $, i.e.,
	\[
	\mathbf{H}^H = \mathbf{U} \mathbf{S} \mathbf{V}^H, \quad
	\mathbf{E} = \mathbf{P} \bm{\Sigma} \mathbf{G}^H,
	\]
	where $\mathbf{S} \in \mathbb{R}^{K \times N}$, $\bm \Sigma \in \mathbb{R}^{N \times L}$ are diagonal matrices and $\mathbf{U} \in \mathbb{C}^{K \times K}$, $\mathbf{V} \in \mathbb{C}^{N \times N}$, $\mathbf{P} \in \mathbb{C}^{N \times N}$,  $\mathbf{G} \in \mathbb{C}^{L \times L}$ are unitary matrices, a relaxed problem of \eqref{eq:channelgain-formula} is obtained as:
	\begin{subequations}\label{eq:SVD}
		\begin{align}
			\max_{\bm \Theta} \quad &\|\mathbf{U}\mathbf{S}\mathbf{V}^H\mathbf \Theta \mathbf{P}\bm \Sigma\mathbf{G}^H\|_F^2\\
			\operatorname{s.t.} \quad \label{con:unitary}&\bm\Theta\bm\Theta^H=\mathbf I_N.
		\end{align}
	\end{subequations}

	Due to the unitary invariance of the Frobenius norm, the objective function of problem \eqref{eq:SVD} is equivalent to $ \|\mathbf{S}\mathbf{V}^H\mathbf \Theta \mathbf{P}\bm \Sigma\|_F^2 $. To gain deeper insight into the channel shaping capability of BD-RIS, we introduce the following definition.
	\begin{definition}(Degree of freedom)\label{definition:DoF}
		The degree of freedom (DoF), also known as multiplexing gain, is the maximum number of independent streams transmitted in parallel over a MIMO channel. The DoF for the compact effective channels $ \mathbf F\triangleq[ \mathbf{f}_1,\cdots, \mathbf{f}_K]\in\mathbb C^{L\times K}$ is defined as  \cite{zhao2024channelshapingusingdiagonal}
		\begin{equation}\label{eq:DoF}
			M=\lim_{\rho \rightarrow \infty} \frac{\log\det\left(\mathbf{I}_L+\rho \mathbf{F}\mathbf{F}^H\right)}{\log \rho}=\min(K,L,N),
		\end{equation}
		where $ \rho $ is the signal-to-noise ratio (SNR).
	\end{definition}
	Based on Definition \ref{definition:DoF}, the matrices $ \mathbf{V} $ and $ \mathbf{P} $ are partitioned as $ \mathbf{V} = [\mathbf{V}_{M},\mathbf{V}_{N-M}] $ and $ \mathbf{P} = [\mathbf{P}_{M},\mathbf{P}_{N-M}] $. Similarly, $ \mathbf{S} $ and $ \bm \Sigma $ are partitioned as $\mathbf{S}=\operatorname{ diag}(\mathbf{S}_{M}, \mathbf{S}_{K-M}) $ and $ \bm \Sigma = \operatorname{diag}(\bm \Sigma_M, \mathbf{\Sigma}_{L-M}) $. Since the DoF of the system is $ M $, both $ \mathbf{S}_{K-M} $ and $ \mathbf{\Sigma}_{L-M} $ are zeros. In this way, we have
	\begin{equation}\label{eq:upper-bound}
		\begin{split}
			\|\mathbf{S}\mathbf{V}^H\bm \Theta \mathbf{P}\bm \Sigma\|_F^2 
			=& \left\| \mathbf S_M\mathbf V_M^H\mathbf\Theta\mathbf P_M\mathbf \Sigma_M \right\|^2_F    \\
			\overset{\underset{\mathrm{(a)}}{}}{\leq}& \left\|\mathbf{S}_{M}\bm \Sigma_M\right\|_F^2,
		\end{split}
	\end{equation}
	where the equality in $\mathrm{(a)}$ holds if and only if
	\begin{equation}\label{eq:optimal-T}
		\mathbf{V}_{M}^H\mathbf \Theta \mathbf{P}_{M} = \bm\Phi,
	\end{equation}
	and $\bm\Phi=\mathrm{diag}([e^{j\phi_1},\cdots,e^{j\phi_M}])$ with $\phi_m\in[0,2\pi), m=1,\cdots,M$. This is a sufficient condition to achieve the theoretic upper bound of problem \eqref{eq:channelgain-formula}. The solution of \eqref{eq:optimal-T} is given by \begin{equation}\label{eq:theta-opt}
		\bm \Theta^\star=  \mathbf{V}_{M} \bm\Phi\mathbf{P}_{M}^H + \mathbf{V}_{N-M} \mathbf X_0\mathbf{P}_{N-M}^H,
	\end{equation}
	where $ \mathbf{X}_0$ is an unitary matrix. Without loss of generality, we set $\phi_m=0$ in the following discussion, as its value does not affect the final objective value. Therefore, the solution in \eqref{eq:theta-opt} can be simplified as
	\begin{equation}\label{eq:theta-symplified}
		\bm \Theta^\star=  \mathbf{V}_{M} \mathbf{P}_{M}^H + \mathbf{V}_{N-M} \mathbf{X}_0\mathbf{P}_{N-M}^H.
	\end{equation}
	This upper bound can provide more insights in channel rearrangement and channel space alignment, quantifying the maximum achievable singular values, power gain, and capacity, revealing BD-RIS’s ability to significantly enhance channel performance compared to diagonal RIS \cite{zhao2024channelshapingusingdiagonal}. However, it is worth noting that, limited by the physical reciprocal property, i.e., $\mathbf \Theta=\mathbf \Theta^T$, all BD-RIS architectures can not generally attain this theoretical upper bound as shown in the following Proposition \ref{pro:fully-con}.
	
	\begin{proposition}\label{pro:fully-con}
		Fully connected RIS can not achieve the performance upper bound in \eqref{eq:upper-bound} when the DoF for the compact effective channels $ \mathbf F$ is larger than 1, i.e., $M>1$. It only achieves the performance upper bound when $ M=1 $. 
	\end{proposition}
	\textit{Proof.} 
	See Appendix \ref{app:proof-fully}.
	$\hfill\blacksquare$
	
	Since fully-connected RIS offers the best performance among BD-RIS structures (feasible set: $\mathcal{B}_{\text{Fu}}\geq \mathcal{B}_{x}, x\in \{\text{Si}, \text{Tr}, {\text{Gr}(G)}, {\text{Fo}(G)}, {\text{St}(Q)},  {\text{Cl}(Q_G,G)}\}$), and it cannot achieve the upper bound in \eqref{eq:upper-bound} when $M>1$, neither can any other BD-RIS architectures.

	\section{ A Novel Structure-Oriented Symmetric and Unitary Projection Framework for BD-RIS}\label{Sec:projection} 
	In this section, we propose a novel structure-oriented symmetric and unitary projection (SOSUP) framework, which generalizes the symmetric unitary projection tailored for fully-connected RIS to accommodate all RIS architectures. Then, we specify two algorithms based on the proposed projection framework to address \eqref{eq:channelgain-formula}. 
	
	In the following, we use the stem-connected RIS as an example to illustrate the fundamentals of the proposed projection operator. It is worth noting that each group within a cluster-connected RIS can be viewed as an individual stem-connected RIS. As a result, the projection operator can be naturally and easily extended to the cluster-connected RIS architecture.
	
	\subsection{The Proposed Structure-Oriented Symmetric Unitary Projector}
	The proposed structure-oriented symmetric unitary projector for stem-connected RIS is detailed in the following definition.
	\begin{definition}
		Let $\mathbf X\in\mathbb C^{N\times N}$ be a given square matrix, the projection operator $\bm\Pi_{\mathcal B_x}(\mathbf{X})$ which projects $ \mathbf{X} $ onto the feasible set $\mathcal B_x$ is defined as
		\begin{subequations}\label{projection problem}
			\begin{align}
				\bm\Pi_{\mathcal B_x}(\mathbf X)\triangleq \arg \min_{\mathbf B}\,\, &\left\|\mathbf X-\mathbf\Theta\right\|^2_F \label{eq:obj-proj}\\
				\operatorname{s.t.}\,\, &\bm \Theta=\left(\mathbf{I}+jZ_0\mathbf{B}\right)^{-1}\left(\mathbf{I}-jZ_0\mathbf{B}\right), \label{non-linear theta}\\
				&\mathbf{B} \in \mathcal{B}_x, \label{eq:gen-stru}
			\end{align}
		\end{subequations}
		where $\|\cdot\|_F$ denotes the Frobenius Norm.
	\end{definition}
	
	
	First, we find the lower bound of the objective function \eqref{eq:obj-proj} and the corresponding sufficient condition to attain this lower bound. Specifically, \eqref{eq:obj-proj} is equivalent to:
	\begin{subequations}
		\begin{align}
			&\left\|\mathbf X-\mathbf\Theta\right\|^2_F\\
			=&
			\left\|\mathbf X-\frac{1}{2}\left(\mathbf X+\mathbf X^T\right)+\frac{1}{2}\left(\mathbf X+\mathbf X^T\right)-\mathbf \Theta\right\|_F^2\\
			=& \left\|\frac{1}{2}\left(\mathbf X+\mathbf X^T\right)-\mathbf \Theta\right\|_F^2\\
			+&2\Re\left\{\mathrm{tr}\left(\frac{1}{2}\left(\mathbf X-\mathbf X^T\right)\left(\frac{1}{2}\left(\mathbf X+\mathbf X^T\right)-\mathbf \Theta\right)\right)\right\}\\
			\overset{(a)}{=}& \left\|\frac{1}{2}\left(\mathbf X+\mathbf X^T\right)-\mathbf \Theta\right\|_F^2\\
			\overset{(b)}{=}& \left\|\mathbf Q\mathbf \Sigma\mathbf Q^T-\mathbf \Theta\right\|_F^2 \label{eq:definite_Q}\\
			\overset{(c)}{=}& \left\|\mathbf Q_R\mathbf \Sigma_R\mathbf Q_R^T-\mathbf \Theta\right\|_F^2\\
			\overset{(d)}{=}& \left\|\mathbf \Sigma_R-\mathbf Q_R^H\mathbf \Theta\mathbf Q_R^*\right\|_F^2,
		\end{align}
	\end{subequations}
	where $(a)$ follows the fact that the inner product of a skew-symmetric matrix and a symmetric matrix is zero, $(b)$ applies the Takagi decomposition \cite{takagi2023} of a certain symmetric matrix $\frac{1}{2}(\mathbf X+\mathbf X^T)=\mathbf Q\mathbf \Sigma\mathbf Q^T $, where $\mathbf Q$ is unitary and $\bm\Sigma$ is diagonal. In $(c)$, $R$ is the rank of $\bm\Sigma$ and $\mathbf Q$ and $\mathbf\Sigma$ are equivalent to $\mathbf Q=[\mathbf Q_R,\mathbf Q_{N-R}]$ and $\mathbf \Sigma=\mathrm{diag}(\mathbf \Sigma_R,\mathbf\Sigma_{N-R})$.
	Since both $\bm\Theta$ and $\mathbf Q$ are unitary matrices, we have
	\begin{equation}\label{eq:lower-bound}
		\left\|\mathbf \Sigma_R-\mathbf Q_R^H\mathbf \Theta\mathbf Q_R^*\right\|_F^2\geq \left\|\mathbf \Sigma_R-\mathbf I_R\right\|_F^2,
	\end{equation}
	where the equality holds if
	\begin{equation}\label{eq:eq-con}
		\mathbf Q_R^H\bm\Theta\mathbf Q_R^*=\mathbf I_R.
	\end{equation}
	Therefore, the right-hand side of inequality \eqref{eq:lower-bound} is a lower bound of \eqref{eq:obj-proj}, it is achieved when \eqref{eq:eq-con} holds. By replacing the objective function \eqref{eq:obj-proj} with $||\bm \Sigma_R-\mathbf{I}_R||^2_F$, we obtain a lower-bound problem of \eqref{projection problem}, and its optimal solution can be directly obtained by solving:
	\begin{equation}\label{linear equation system for projection}
		\eqref{eq:eq-con},\eqref{non-linear theta} \; \text{and} \; \eqref{eq:gen-stru}.
	\end{equation}

	Next, we proposed a structure-oriented symmetric unitary projection to solve problem \eqref{linear equation system for projection}. Specifically, by replacing $\bm \Theta$ in \eqref{eq:eq-con} with \eqref{non-linear theta}, we obtain
	\begin{equation}\label{eq:BCD}
		\mathbf{B}\mathbf{C}=\mathbf{D},
	\end{equation}
	where $\mathbf{C}=jZ_0(\mathbf{Q}_R^*+\mathbf{Q}_R)$, and $\mathbf{D}=\mathbf{Q}_R-\mathbf{Q}_R^*$. Given that $\mathbf{B}$ is a real-value matrix, we further equivelently transform equation \eqref{eq:BCD} into its real and imaginary parts, yielding:
	\begin{subequations}
		\begin{align}
			\mathbf{B} \Re\{\mathbf{C}\} = \Re \{\mathbf{D}\},\label{eq:pure-image}\\
			\mathbf{B} \Im\{\mathbf{C}\} = \Im \{\mathbf{D}\},\label{eq:re-im}
		\end{align}	
	\end{subequations} 
	where $\Re(\cdot)$ and $\Im(\cdot)$ denote the real and imaginary parts of a complex matrix, respectively. Since both $\mathbf{C}$ and $\mathbf{D}$ are purely imaginary, \eqref{eq:pure-image} can be omitted.
	
	By employing the standard vectorization operator $\operatorname{vec}(\mathbf{X})$ (which simply stacks the columns of $ \mathbf{B} $ on top of one another) to both sides of \eqref{eq:re-im}, \eqref{eq:re-im} is equivalent to:
	\begin{equation}\label{eq:re-im-vec}
		\left(\Im\{\mathbf{C}\}^T \otimes \mathbf{I}_N\right) \operatorname{vec}(\mathbf{B}) = \operatorname{vec}(\Im\{\mathbf{D}\}),
	\end{equation}
	where $\otimes$ denotes the Kronecker product.

	Before delving deeper into \eqref{eq:re-im-vec}, we further define a novel vectorization operator to capture the inherent structure constraint \eqref{eq:B-stru} in matrix $\mathbf{B}$.
	\begin{definition}[Independent Vectorization]\label{definition:vec} For any matrix $\mathbf{B}$, the operator $\operatorname{vec_i}(\mathbf{B})$ extracts all independent variables in $\mathbf{B}$ and arranges them into a column vector. 
	\end{definition}

	\par  For the proposed stem-connected RIS architecture described in Section~\ref{sec:stem-con}, matrix $\mathbf{B}$ contains $2QN + N - Q(Q + 1)$ non-zero elements, as illustrated in Fig.~\ref{fig:arrow}. Since $\mathbf{B}$ is symmetric in the stem-connected architecture, i.e., $\mathbf{B} = \mathbf{B}^T$, the number of independent variables is reduced to $QN + N - \frac{Q(Q + 1)}{2}$. 
	Hence, the independent vectorization of $\mathbf{B}$ is defined as
	\begin{equation} 
		\begin{split}
			\operatorname{vec_i}(\mathbf{B}) &= \left[[\mathbf{B}]_{1,1}, \ldots, [\mathbf{B}]_{1,N}, \ldots, [\mathbf{B}]_{Q,Q}, \ldots, [\mathbf{B}]_{Q,N},\right.\\ &\left.[\mathbf{B}]_{Q+1,Q+1}, \ldots, [\mathbf{B}]_{N,N}\right]^T\in \mathbb{R}^{\left( QN+N-\frac{Q(Q+1)}{2} \right) \times 1}.
		\end{split}
	\end{equation}
	It is worth noting that the independent vectorization $\operatorname{vec_i}(\mathbf{B})$ is a linear transformation of the conventional vectorization operator $\operatorname{vec}(\mathbf{B})$ by removing all zero and repeated elements.

	Then, let $\mathbf{R} \in \mathbb{R}^{N^2 \times (QN + N - \frac{Q(Q+1)}{2})}$, and define $\widetilde{\mathbf{b}} \triangleq \operatorname{vec}(\mathbf{B})$ and $\mathbf{b} \triangleq \operatorname{vec_i}(\mathbf{B})$, we have 
	\begin{equation}\label{eq:nullspace}
		\widetilde{\mathbf{b}}=\mathbf{R}\mathbf{b},
	\end{equation}
	where $\mathbf{R}$ maps  $\mathbf{b}$ to  $	\widetilde{\mathbf{b}}$ so as to satisfy the  stem-connected RIS constraints $\mathbf{B} \in \mathcal{B}_{\text{St}(Q)}$. To mathematically specify $ \mathbf{R} $, we first introduce the following indicator matrix $ \mathbf{A}_B $ and non-zero counting matrix  $ \mathbf{C}_B $ for $ \mathbf{B} $.

	\begin{definition}(Indicator matrix)\label{def:indicator}
		Given a symmetric matrix $ \mathbf{B} \in \mathbb{R}^{N \times N}$, we define an indicator matrix $ \mathbf{A}_B $ of $\mathbf{B}$ to identify the positions of non-zero elements in $ \mathbf{B} $. Specifically, each element of $ \mathbf{A}_B \in \mathbb{R}^{N \times N} $ is defined as:
		\begin{equation}\label{eq:indicator-B}
			[\mathbf{A}_B]_{i,j}= \left\{\begin{matrix}
				1,	& \text{if} \; [\mathbf{B}]_{i,j} \neq 0 \\
				0,	& \text{if} \; [\mathbf{B}]_{i,j}=0 \\
			\end{matrix}\right..
		\end{equation}
	\end{definition}

	\begin{definition}(Non-zero counting matrix)\label{def:count}
		Given an indicator matrix $ \mathbf{A}_B $ for a symmetric matrix $ \mathbf{B} \in \mathbb{R}^{N \times N}$, we define a transformation to generate a matrix \( \mathbf{C}_B \) that reflects the cumulative counting of the upper triangular part of $ \mathbf{A}_B $. $ \mathbf{C}_B $ is obtained by the following four steps:
		\begin{enumerate}
			\item Initialize $ \mathbf{A}_B $ as an $ N \times N $ zero matrix.
			\item Starting from $ [\mathbf{A}_B]_{1,1} $, traverse the upper triangular part of $ \mathbf{A}_B $ (including the main diagonal, i.e., $ i \leq j $) in row-major order.
			\item Use a counter $ c $, initialized to 0. For each position $ (i,j) $:
			\begin{itemize}
				\item If $ [\mathbf{A}_B]_{i,j} = 1 $, then $ c = c + 1 $, and assign $ [\mathbf{C}_B]_{i,j}=c $;
				\item If $ [\mathbf{A}_B]_{i,j} = 0 $, then $ [\mathbf{C}_B]_{i,j} $ retains the count value from the previous position (or 0 if no previous position exists).
			\end{itemize}
			\item Fill the lower triangular part ($ i > j $) symmetrically: $ [\mathbf{C}_B]_{i,j} = [\mathbf{C}_B]_{j,i} $.
		\end{enumerate}	
	\end{definition}
	
	Based on Definition~\ref{def:count}, we define the transformation matrix $\mathbf{R} \in \mathbb{R}^{N^2 \times \left(QN + N - \frac{Q(Q+1)}{2}\right)}$ as a row-wise stacking of vectors, i.e., $\mathbf{R} = [\mathbf{R}_1^T, \cdots, \mathbf{R}_m^T, \cdots, \mathbf{R}_{N^2}^T]^T$, where each row vector $\mathbf{R}_m$ consists of only zeros and ones. Define ${\mathbf{a}}_B \triangleq \operatorname{vec}(\mathbf{A}_B)$ according to Definition \ref{def:indicator}, the structure of $ \mathbf{R}_m $ in $\mathbf{R}$ is detailed as follows:
	\begin{equation}\label{eq:R_m}
		\mathbf{R}_m= \left\{\begin{matrix}
			\mathbf{e}_n^T,	& \text{if} \; {\mathbf{a}}_B[m]=1 \\
			\mathbf{0},	& \text{if} \; {\mathbf{a}}_B[m]=0 \\
		\end{matrix}\right.,
	\end{equation}
	where $ \mathbf{R}_m \in \mathbb{R}^{1 \times (QN + N - \frac{Q(Q+1)}{2})} $, $ {\mathbf{a}}_B[m]=[\mathbf{A}_B]_{i,j}$, $ n = [\mathbf{C}_B]_{i,j} $, and $ \mathbf{e}_n $ is the $ n $-th column of an identity matrix.

	Matrix $ \mathbf{R} $ is crucial as it linearly maps $ \operatorname{vec}_i(\mathbf{B}) $ to $ \operatorname{vec}(\mathbf{B}) $ while imposing BD-RIS structural constraints. Its mathematical form varies across architectures, reflecting their unique connectivity patterns.
	For better intuition, a toy example illustrating $\mathbf{R}$ is provided in Appendix \ref{app:R-example}.

	Substituting \eqref{eq:nullspace} into \eqref{eq:re-im-vec}, problem \eqref{linear equation system for projection} is equivalently transformed as:
	\begin{equation}\label{eq:Abz}
		\mathbf{A}\mathbf{b}=\mathbf{z},
	\end{equation}
	where $ \mathbf{A}= \left(\Im\{\mathbf{C}\}^T  \otimes \mathbf{I}\right) \mathbf{R} \in \mathbb{R}^{RN \times \left(\frac{2N-1}{2}Q-\frac{Q^2}{2}+N\right)}$, $ \mathbf{z}=\operatorname{vec}(\Im \{\mathbf{D}\}) \in \mathbb{R}^{RN\times 1} $. We then employ the least square method to solve \eqref{eq:Abz}, leading to the solution:
	\begin{equation}\label{eq:approximate-solution}
		\mathbf{b}^* = \left(\mathbf{A}^T\mathbf{A}\right)^{-1}\mathbf{A}^T\mathbf{z}.
	\end{equation}
	Once we obtain $\mathbf{b}^*$, which contains the independent variables in $\mathbf{B}$, we can employ the relationship in~\eqref{eq:indicator-B} to recover $\mathbf{B}$ with specific structure, i.e., $\mathbf{B} = \mathrm{vec}^{-1}_{N,N}(\mathbf{R}\mathbf{b}^*)$.  The operation $\mathbf{A}=\operatorname{vec}_{N, N}^{-1}(\mathbf{a})$ reshapes a vector $\mathbf{a} \in \mathbb{R}^{N^2 \times 1}$ into an $N \times N$ matrix $\mathbf{A} \in \mathbb{R}^{N \times N}$ by arranging the $N^2$ elements of $\mathbf{a}$ column-wise, such that the first $N$ elements form the first column, the next $N$ elements form the second column, and so on. 
	
	\setlength{\textfloatsep}{7pt}	
	\begin{algorithm}[t!]
		\caption{Proposed structure-oriented symmetric unitary projection (SOSUP) for problem \eqref{projection problem}}
		\label{alg:low-complexity}
		\textbf{Input:} A square matrix $ \mathbf X $\;
		{
			Obtain $ \mathbf{Q} $ and $ \mathbf{Q}^* $ through Takagi decomposition of $\frac{1}{2}(\mathbf{X}+\mathbf{X}^T)$;\\
			Obtain $ \mathbf{Q}_{R} $ and $ \mathbf{Q}_{R}^* $ defined in \eqref{eq:definite_Q};\\
			Obtain $ \mathbf{R} $, $ \mathbf{C} $ and $ \mathbf{D} $ defined in \eqref{eq:nullspace} and \eqref{eq:BCD};\\
			Obtain $ \mathbf{A} $ and $ \mathbf{z} $ defined in \eqref{eq:Abz};\\
			Obtain $ \mathbf{b}^* $ by \eqref{eq:approximate-solution};\\
			Recover $\mathbf{B}$ through $\mathbf{B} = \mathrm{vec}^{-1}_{N,N}(\mathbf{R}\mathbf{b}^*)$\;
		}
	\end{algorithm}
	
	The details of our proposed SOSUP is illustrated in Algorithm \ref{alg:low-complexity}. Since \eqref{eq:Abz} is equivalent to \eqref{linear equation system for projection}, which represents the theoretical lower bound of \eqref{eq:lower-bound}, the solution generated by Algorithm \ref{alg:low-complexity} is guaranteed to approach this fundamental performance limit. 
	The computational complexity of Algorithm \ref{alg:low-complexity} is dominated by Step 6, which has an order of $ \mathcal{O}(Q^3N^3) $.
	
	\subsection{Proposed Algorithms for the Channel Gain Maximization Problem \eqref{eq:channelgain-formula}}\label{sec:channel-gain-sol}
	Building on the proposed SOSUP, we here propose two algorithms to solve the sum channel gain maximization problem \eqref{eq:channelgain-formula}:
	\begin{itemize}

		\item \textbf{UB-based SOSUP}: Although the optimal solution to the upper-bound problem in \eqref{eq:theta-symplified} is generally infeasible under the constraints of \eqref{eq:channelgain-formula}, we can project this solution onto the feasible set defined by $\mathcal{B}_x$. Inspired by this idea, we propose a low-complexity method named \textit{UB-based SOSUP}, which stands for upper-bound-initiated structure-oriented symmetric unitary projection. This method leverages a closed-form solution to efficiently solve the sum channel gain maximization problem. Specifically, the projection is given by
		\begin{align}\label{eq:LS-solution}
			\mathbf{B} = \bm\Pi_{\mathcal{B}_x}(\mathbf{V}_M \mathbf{P}_M^H),
		\end{align}
		where $\mathbf{V}_M \mathbf{P}_M^H$ is obtained by setting $\mathbf{X}_0 = 0$ in \eqref{eq:theta-symplified}. The main difference among various BD-RIS architectures lies in the constraint set $\mathcal{B}_x$, which is determined by the design of the transformation matrix $\mathbf{R}$. Therefore, the proposed UB-based SOSUP method is broadly applicable to all RIS architectures by appropriately tailoring $\mathbf{R}$ for each case.

		\item \textbf{SOSUP-based quasi-Newton}: Based on the independent vectorization defined in Definition \ref{definition:vec}, problem \eqref{eq:channelgain-formula} is equivalently reformulated as an unconstrained optimization problem as follows:
		\begin{equation}\label{eq:pro-quasi}
			\max_{\mathbf{b}} \quad \|\mathbf H^H \left(\mathbf{I}+jZ_0\mathbf{B}\right)^{-1}\left(\mathbf{I}-jZ_0\mathbf{B}\right)\mathbf E\|_F^2 
		\end{equation}
		where $ \mathbf{b}=\operatorname{vec}_i(\mathbf{B}) $. \eqref{eq:pro-quasi} can be directly solved using the quasi-Newton method. However, the quasi-Newton method is prone to getting trapped in local optima. Noting that the proposed \textit{UB-based SOSUP} method is an efficient closed-form solution, it can efficiently provide a high-quality initialization for quasi-Newton method. For convenience, we denote the quasi-Newton method initialized by \eqref{eq:LS-solution} as the \textit{SOSUP-based quasi-Newton} method. 
	\end{itemize}
	
	\subsection{Extension to the Joint Active and Passive Beamforming Optimization}\label{Sec:WSR}
	
	In this subsection, we further clarify how to extend the SOSUP algorithm to tackle the following joint active and passive beamforming optimization problem under the total transmit power constraint at the BS, i.e., $ \|\mathbf W\|^2_F \leq  P_t $. Following the system model specified in Section \ref{sec:sys-mod}, the achievable rate at user $ k $ to decode the intended symbol is given by
	\begin{equation}\label{eq:rate}
		\begin{aligned}
			R_k= \log \left(1+ \frac{|\mathbf f_k^H\mathbf w_k|^2}{\sum_{j=1,j\neq k}^K |\mathbf {f}_k^H \mathbf w_j|^2+\sigma_k^2  }  \right).
		\end{aligned}
	\end{equation}
	Consider a general utility function $  f(R_1, \cdots, R_K) $, where $ f(\cdot) $ can be a classical performance metrics such as:
	\begin{itemize}
		\item Weighted sum rate (WSR): $ \sum_{k=1}^K \delta_k R_k $,
		where $\delta_k$ is the weight of user $k$.
		\item Max-min fairness (MMF) rate: $ \min_{k \in \mathcal{K}} \{R_k\}. $
		\item Energy efficiency (EE): $ \frac{\sum_{k=1}^K \delta_k R_k}{\frac{1}{\eta}\|\mathbf W\|^2_F+P_{\text{cir}}} $,
		where $ \eta $ is the power amplifier coefficient and $ P_{\text{cir}} $ is the circuit power.
	\end{itemize}
	The corresponding optimization problem can be formulated as:
	\begin{subequations}\label{eq:utility-pro}
		\begin{align}
			\max_{\bm\Theta, \mathbf W}\,\, &  f(R_1, \cdots, R_K)\\
			\operatorname{s.t.} \quad &\bm \Theta=\left(\mathbf{I}+jZ_0\mathbf{B}\right)^{-1}\left(\mathbf{I}-jZ_0\mathbf{B}\right), \\
			&\mathbf{B} \in \mathcal{B}_x,\\
			&\|\mathbf W\|^2_F \leq P_t.
		\end{align}
	\end{subequations}

	Since both the objective function and the passive beamforming constraint are non-convex, problem \eqref{eq:utility-pro} is highly challenging to solve. In general, system performance is maximized by jointly designing the active and passive beamforming matrices. An effective approach to solving problem~\eqref{eq:utility-pro} with the objective of maximizing WSR is to first ignore the structural constraints imposed by the BD-RIS architecture, solve the relaxed problem, and then project the solution onto the feasible set that satisfies the BD-RIS structural constraints, as suggested in~\cite{hongyu2023,hongyu2024,hongyu2023jasc,yuyuan2023,fang2023low,202501joint}. However, the existing projection methods proposed in~\cite{hongyu2023,hongyu2024,hongyu2023jasc,yuyuan2023,fang2023low,202501joint} are specifically tailored for fully-connected RIS architectures, where the solution of the relaxed problem is projected onto the set of matrices that satisfy the symmetric unitary constraint $\mathcal{B}_{\text{fu}}$. These methods are not applicable to other BD-RIS architectures that involve additional structural constraints, such as the proposed stem-connected RIS characterized by~\eqref{eq:B-arrow}.
	In contrast, the method proposed in~\cite{wu2024-ppADMM} is designed to directly address problem~\eqref{eq:utility-pro}. It decomposes the original problem into two subproblems corresponding to active and passive beamforming, which are solved alternately. However, this alternating approach results in relatively high average CPU complexity.

	Inspired by our proposed \textit{UB-based SOSUP} and \textit{SOSUP-based quasi-Newton} algorithms for addressing the channel gain maximization problem in Section \ref{sec:channel-gain}, we propose a highly efficient two-stage beamforming design algorithm to address \eqref{eq:utility-pro}. Specifically, in the first stage, we directly design the passive beamforming matrix to enhance the effective channel gain by solving problem \eqref{eq:channelgain-formula}; in the second stage, we design the active beamforming matrix to maximize the system utility with the fixed passive beamforming matrix obtained in the first stage. The detailed description of the two-stage algorithm is as follows:  
	
	\begin{itemize}
		\item \textit{Stage 1:} Design the passive beamforming matrix $\bm{\Theta}$ by solving sum channel gain maximization problem \eqref{eq:channelgain-formula} in Section \ref{sec:channel-gain} via the two methods illustrated in Section \ref{sec:channel-gain-sol}.
		\item \textit{Stage 2:} With the designed $\bm{\Theta}$ from the first stage, solve the active beamforming problem using classical beamforming optimization algorithms such as fractional programming (FP) algorithm~\cite{shen2018fractional,202501joint}, successive convex approxiamtion (SCA)~\cite{2025energy}, cyclic maximization (CM)-based method~\cite{fang2023multi-group}.
	\end{itemize}
	
	
	\vspace{-4.5mm}
	\section{Simulation Results}\label{Sec:simulation}

	In this section, we evaluate the performance of the two proposed novel BD-RIS architectures, namely the stem-connected RIS and the cluster-connected RIS. We also assess the performance of the proposed SOSUP algorithms with respect to both the sum channel gain maximization problem and the WSR maximization problem for the proposed BD-RIS architectures.
	
	\par Following \cite{fu2019}, we consider the following channel model for $ \mathbf{E} $ and $ \mathbf{h}_k $:
	\begin{equation}
		\quad \mathbf{E} = \sqrt{P(d_r)}\bm \Gamma, 	\mathbf{h}_k = \sqrt{P(d_k)}\mathbf{r},
	\end{equation}
	where $\bm \Gamma \sim \mathcal{CN}(0,\mathbf{I})$ and  $\mathbf{r} \sim \mathcal{CN}(0,\mathbf{I})$. The distance between the BS and the BD-RIS, denoted by $d_r$, and the distance between the BD-RIS and the user $ k $, denoted by $d_k$, are set to $50\sqrt{2}$ and $ 50\sqrt{5}$ meters, respectively. The reference path loss at a link distance of $d_0 = 1$ meter is given as $L_0 = -30$ dB, and the path loss model is expressed as $P(d) = L_0 \left(d/d_0\right)^{-\alpha}$, where $\alpha$ is the path loss exponent. Specifically, $\alpha$ is set to $ 2 $ for the BS–RIS link and $ 2.2 $ for the RIS–user link. The small-scale fading is modeled using Rayleigh fading. All simulation results are averaged over 100 independent channel realizations.

	\subsection{Simulation for the proposed BD-RIS architectures}

	%
	%
	%
	
	\begin{figure}[t!]
		\centering
		\begin{subfigure}{1\linewidth}
			\centering
			\includegraphics[width=0.7\linewidth]{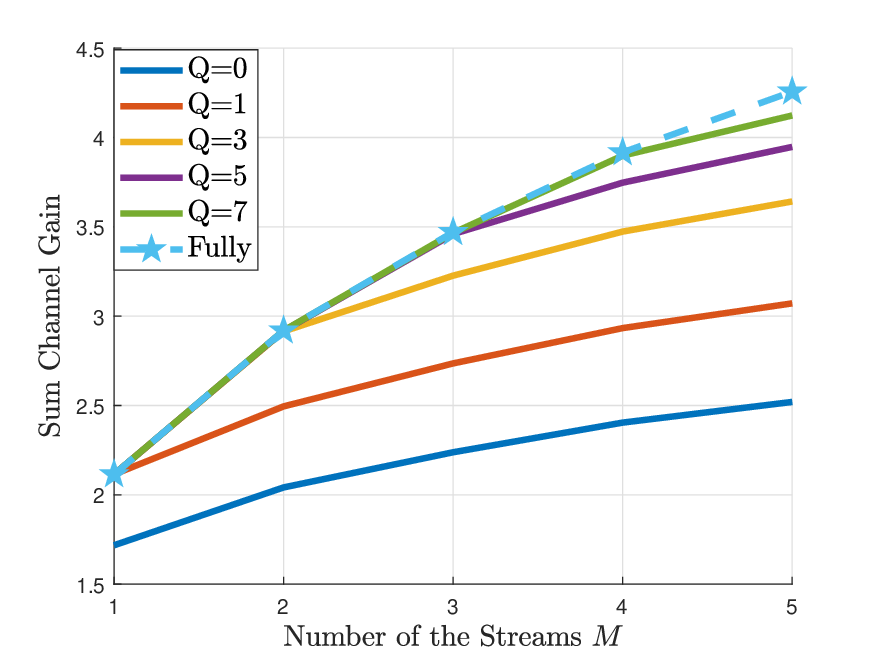}
			\caption{Channel gain versus the number of streams.}
			\label{fig:KL_channel_gain}
		\end{subfigure}
		\vspace{1mm}
		
		\begin{subfigure}{1\linewidth}
			\centering 
			\includegraphics[width=0.7\linewidth]{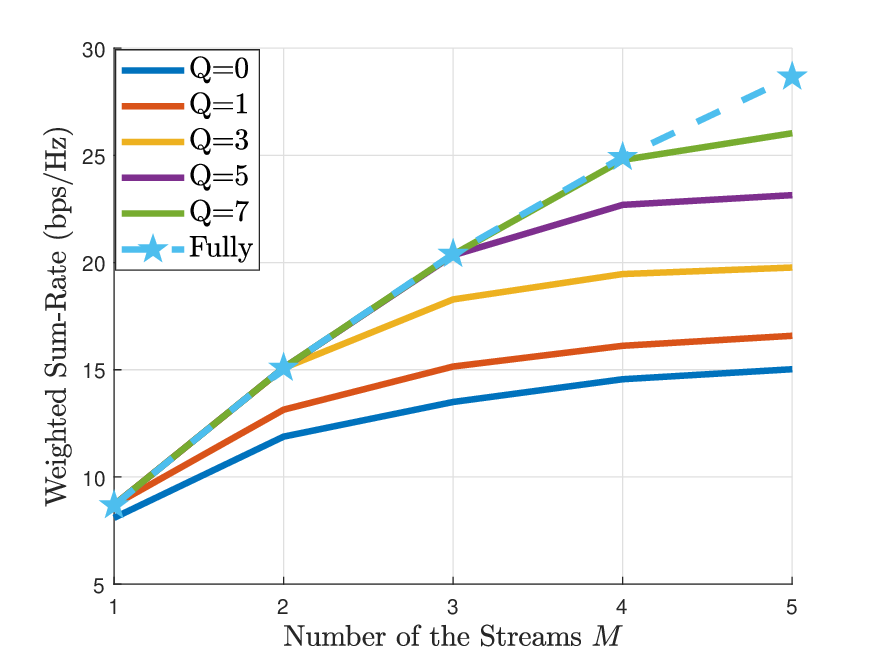}
			\caption{WSR versus the number of streams.}
			\label{fig:KL_WSR}
		\end{subfigure}
		\caption{Performance versus the number of streams when $L = 5$ and $N=64$.}
		\label{fig:L-varyK}
		\vspace{-4mm}
	\end{figure}

	In Fig.~\ref{fig:L-varyK}, we illustrate the system performance with respect to the number of data streams $M$ when $L=5$ and $N=64$. According to Definition~\ref{definition:DoF}, the DoF satisfies $M = \min(K, L, N) = K$ in this setting.  Fig.~\ref{fig:KL_channel_gain} presents the sum channel gain versus $M$, while Fig.~\ref{fig:KL_WSR} shows the WSR performance versus $M$. As observed in both figures, the system performance improves with the increase of $M$. Notably, the performance of the stem-connected RIS approaches that of the fully-connected RIS when $Q = 2M - 1$, which empirically validates the theoretical result established in~\cite{wu2025band}. 

	\begin{figure}[t!]
		\centering
		\begin{subfigure}{1\linewidth}
			\centering
			\includegraphics[width=0.7\linewidth]{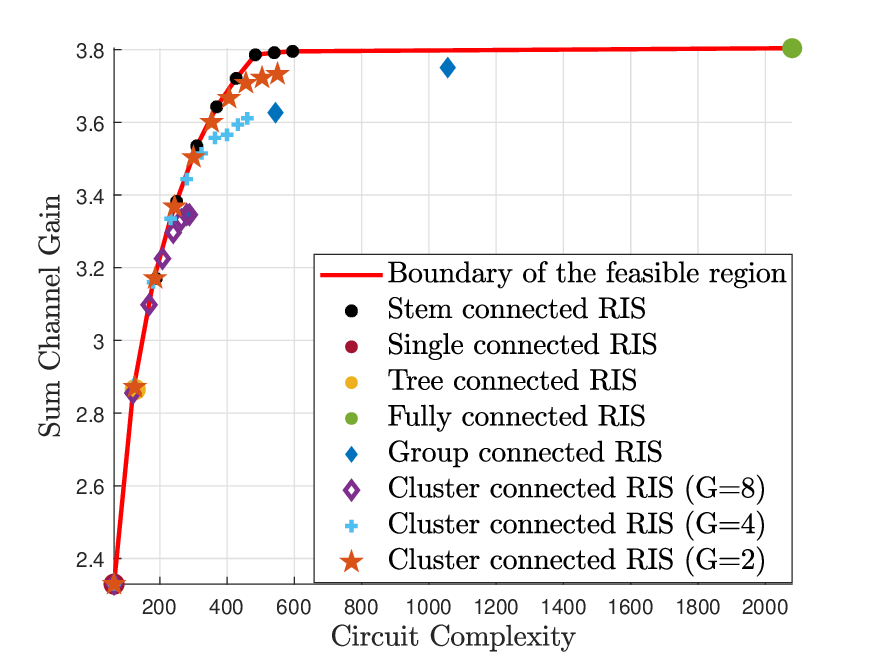}
			\caption{Sum channel gain versus the circuit complexity for BD-RISs.}
			\label{fig:channel-gain-pareto}
		\end{subfigure}
		\vspace{1mm}
		
		\begin{subfigure}{1\linewidth}
			\centering 
			\includegraphics[width=0.7\linewidth]{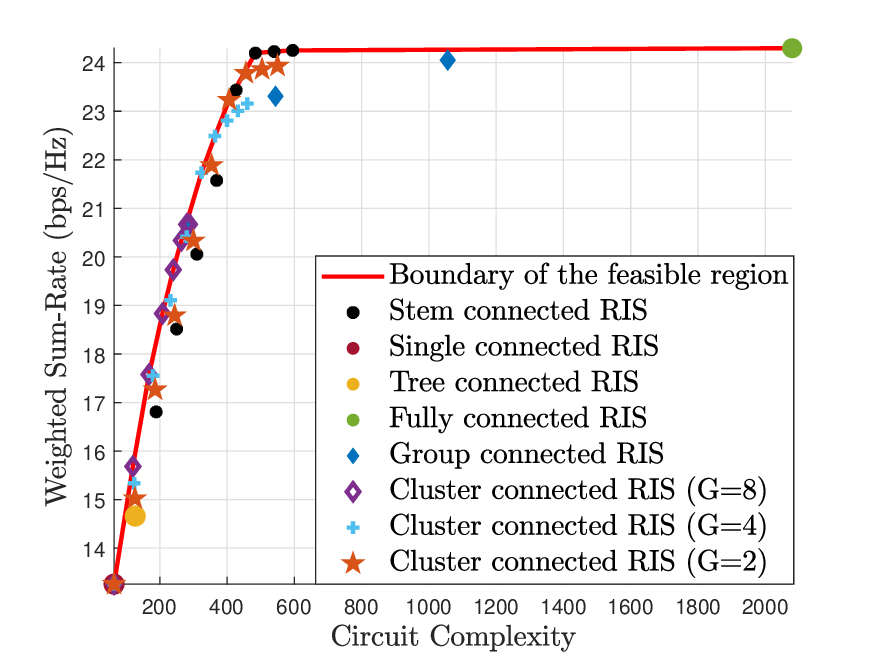}
			\caption{WSR versus the circuit complexity for BD-RISs.}
			\label{fig:sum-rate-pareto}
		\end{subfigure}
		\caption{Feasible region for the performance-complexity trade-off achieved by BD-RISs when $L = K = 4$.}
		\label{fig:pareto}
	\end{figure}

	Fig.~\ref{fig:pareto} illustrates the performance–complexity trade-off region achieved by different BD-RIS architectures with $L=K=4$. Specifically, Fig.~\ref{fig:channel-gain-pareto} shows the sum channel gain versus circuit complexity, while Fig.~\ref{fig:sum-rate-pareto} presents the WSR versus circuit complexity. For example, the sum channel gain of the stem-connected RIS with $Q=7$ is higher than that of the group-connected RIS with $G=4$, while its circuit complexity is lower. The WSR of the cluster-connected RIS with $G=8, Q_G=3$ is higher than that of the stem-connected RIS with $Q=3$, while its circuit complexity is much lower. The proposed stem-connected and cluster-connected RIS architectures explicitly enlarge the performance–complexity trade-off region, clearly outperforming existing BD-RIS structures. In the MU-MISO case, the performance gap between the cluster-connected RIS and the stem-connected RIS can be further enlarged by dynamically tuning the number of RIS elements in different groups of the cluster-connected RIS.

	\subsection{Simulation for the proposed low-complexity algorithms}
	
	The following beamforming design algorithms are compared in this section for different RIS architectures:
	\begin{itemize}
		\item \textit{UB-based SOSUP}: Proposed in Section~\ref{sec:channel-gain-sol} for channel gain maximization; \textit{TS UB-based SOSUP}: Two-stage version in Section~\ref{Sec:WSR} for WSR maximization.
		\item \textit{SOSUP-based quasi-Newton}: Proposed in Section~\ref{sec:channel-gain-sol} for channel gain maximization; \textit{TS SOSUP-based quasi-Newton}: Two-stage variant in Section~\ref{Sec:WSR} for WSR maximization.
		\item \textit{TS SOSUP}: A heuristic two-stage scheme for WSR maximization using $\mathbf{H} \mathbf{I}_{K \times N} \mathbf{E}^H$ as the SOSUP initializer.
		\item \textit{PSLA}: Baseline scheme proposed in~\cite{202501joint} for fully-connected RIS only. 
		\item \textit{LS}: Baseline scheme applied to channel gain maximization in ~\cite{qstem}; \textit{TS LS}: Two-stage variant  for WSR maximization.
		\item \textit{pp-ADMM}: Baseline method in ~\cite{wu2024-ppADMM}.
	\end{itemize}
		\begin{figure}[t!]
		\centering
		\begin{subfigure}{1\linewidth}
			\centering
			\includegraphics[width=0.7\linewidth]{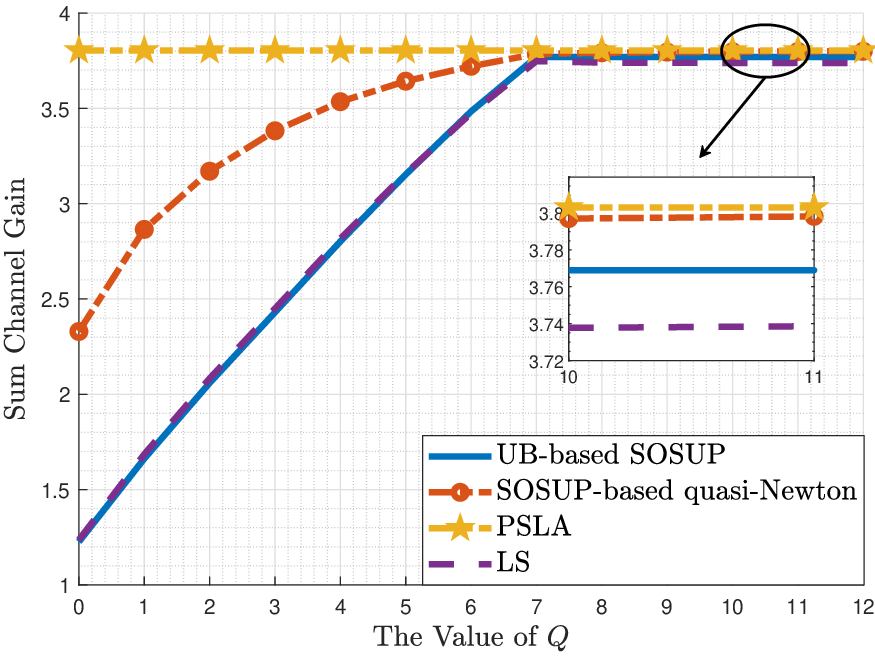}
			\caption{Channel gain versus the value of
				$Q$ in $\mathbf{B}$.}
			\label{fig:YQ-sum-channel-gain}
		\end{subfigure}
		
		\begin{subfigure}{1\linewidth}
			\centering 
			\includegraphics[width=0.7\linewidth]{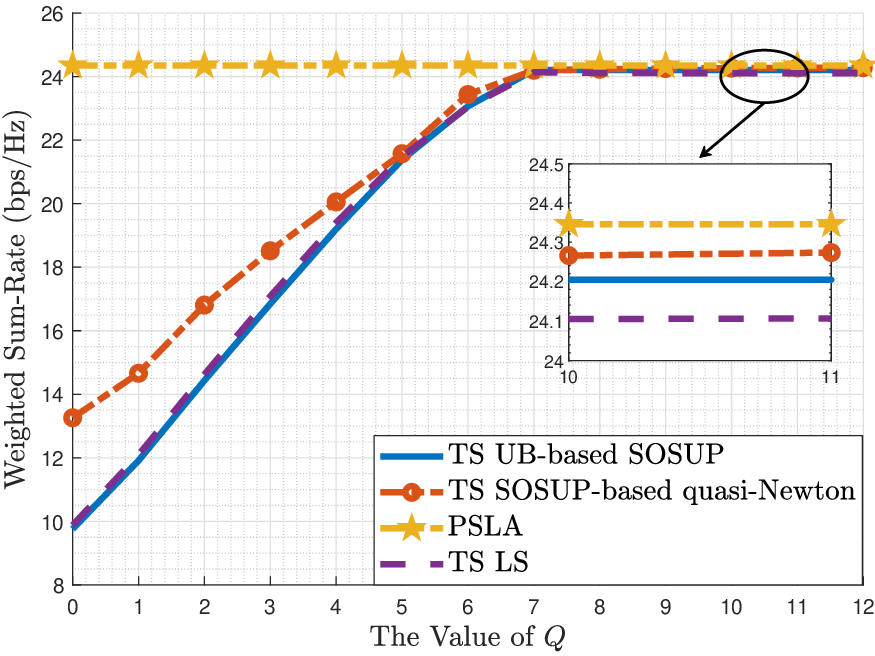}
			\caption{WSR versus the value of
				$Q$ in $\mathbf{B}$.}
			\label{fig:YQ-sum-rate}
		\end{subfigure}
		\caption{Performance versus the value of $Q$ when $L = K = 4$ and $N = 64$.}
		\label{fig:performance-varyQ}
	\end{figure}

	The system performance of the proposed stem-connected RIS with respect to the value of $Q$ in the susceptance matrix $\mathbf{B}$, under the setting $L=K=4$ and $N=64$, is illustrated in Fig.~\ref{fig:performance-varyQ}. Specifically, Fig.~\ref{fig:YQ-sum-channel-gain} and Fig.~\ref{fig:YQ-sum-rate} show the sum channel gain and WSR versus the value of $Q$, respectively. The algorithms based on the proposed SOSUP framework consistently achieve equal or better performance compared to the LS-based method proposed in~\cite{qstem}. In particular, the proposed SOSUP-based quasi-Newton method demonstrates significantly better performance than the LS method when $Q$ is small.

\begin{figure}[t!]
		\centering
		\begin{subfigure}{1\linewidth}
			\centering
			\includegraphics[width=0.7\linewidth]{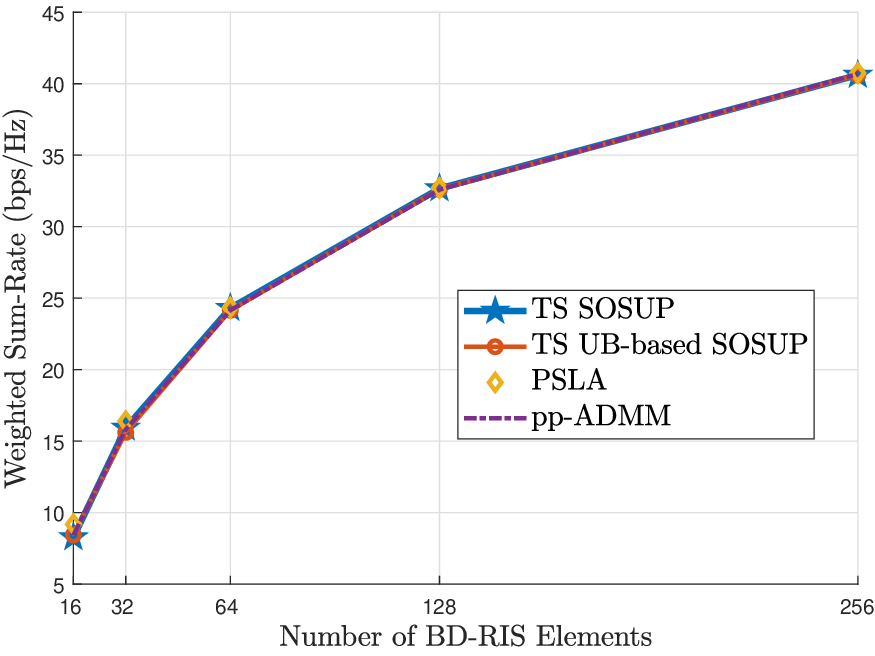}
			\caption{WSR versus the number of BD-RIS elements.}
			\label{fig:WSR-differ-algorithms}
		\end{subfigure}
		\vspace{1mm}
		
		\begin{subfigure}{1\linewidth}
			\centering 
			\includegraphics[width=0.7\linewidth]{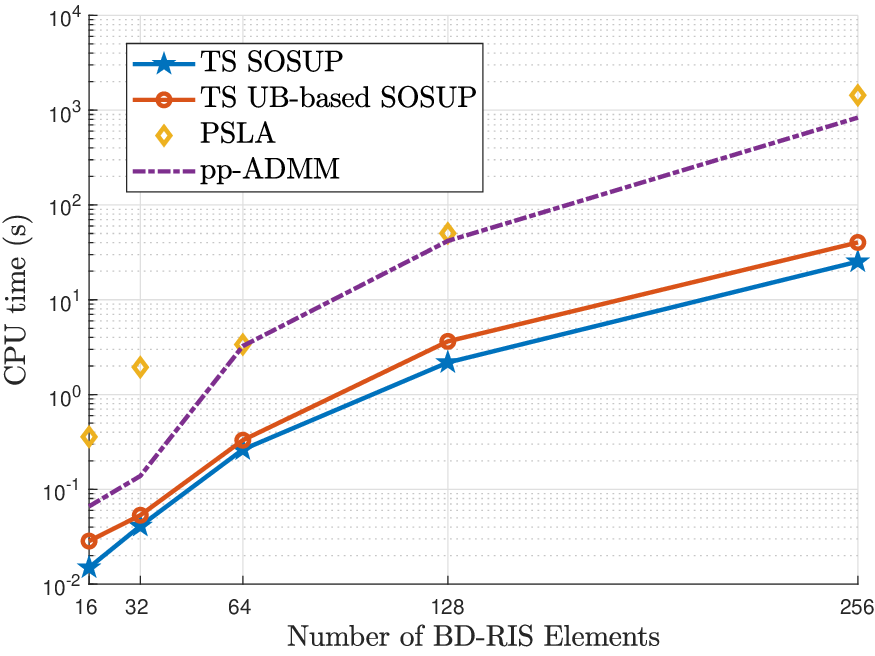}
			\caption{Average CPU time versus the number of BD-RIS elements.}
			\label{fig:WSR-T-differ-algorithms}
		\end{subfigure}
		\caption{Performance versus the number of BD-RIS elements when $L = K = 4$.}
		\label{fig:algs-varyN}
	\end{figure}

	Fig.~\ref{fig:algs-varyN} shows the WSR performance and average CPU time of the proposed SOSUP-based algorithms with varying numbers of BD-RIS elements. Except for PSLA, which applies only to fully connected RIS, all other algorithms are evaluated at $Q=7$, where the stem-connected RIS matches the performance of the fully connected RIS. Fig.~\ref{fig:WSR-differ-algorithms} shows that both proposed two-stage algorithms achieve WSR performance comparable to PSLA and pp-ADMM. Fig.~\ref{fig:WSR-T-differ-algorithms} further shows that TS UB-based SOSUP and TS SOSUP consume only 13.73\% and 21.05\% of the CPU time required by pp-ADMM, respectively. This demonstrates the superior efficiency of the proposed methods in reducing computational complexity without compromising system performance.

	\section{Conclusion}\label{Sec:conclusion}
	
	In this paper, we propose two novel BD-RIS architectures, named stem-connected RIS and cluster-connected RIS. The stem-connected RIS is the simplest BD-RIS architecture that can achieve optimal RIS performance when the number of stems equals to or is larger than the number of data streams. The cluster-connected RIS is another BD-RIS architecture that achieves a flexible and enhanced performance-complexity trade-off between single-connected RIS and fully-connected RIS. Moreover, the proposed cluster-connected RIS serves as a generalized architecture that encompasses all existing BD-RIS architectures as special cases. For the proposed stem-connected RIS and cluster-connected RIS, we introduce an efficient structure-oriented symmetric unitary projection method to calculate the scattering matrix. 
	We further explore the proposed structure-oriented symmetric unitary projection to address the sum channel gain maximization and other utility maximization problems.
	Extensive simulations are conducted to validate the effectiveness of the proposed BD-RIS architectures and the structure-oriented symmetric unitary projection method. From the architecture perspective, the proposed stem-connected RIS achieves the performance of the fully-connected RIS while significantly reducing circuit complexity. In addition, the proposed cluster-connected RIS expands the performance–complexity trade-off region. From an algorithmic perspective, the proposed projection method achieves comparable system performance to existing baselines while substantially reducing average CPU time. 



	\begin{appendices}
		\section{Example of Transformation Matrix}\label{app:R-example}
		Here we provide a toy example of the transformation matrix $\mathbf{R}$ when a stem-connected RIS is considered. Assume $ N=3, Q=1 $, and $ \mathbf{B} $ is given as follows
		\begin{equation}
			\mathbf{B}=\begin{bmatrix}
				[\mathbf{B}]_{1,1} & [\mathbf{B}]_{1,2} & [\mathbf{B}]_{1,3}  \\
				[\mathbf{B}]_{1,2}& [\mathbf{B}]_{2,2} & 0  \\
				[\mathbf{B}]_{1,3}&0 & [\mathbf{B}]_{3,3} 
			\end{bmatrix}.
		\end{equation}
		The corresponding indicator matrix $ \mathbf{A}_B $ and non-zero counting matrix $ \mathbf{C}_B $ according to Definition \ref{def:indicator} and \ref{def:count} are given as follows:
		\begin{equation}
			\mathbf{A}_B=\begin{bmatrix}
				1 & 1 & 1  \\
				1& 1 & 0  \\
				1&0 & 1 
			\end{bmatrix}, \mathbf{C}_B=\begin{bmatrix}
				1 & 2 & 3  \\
				2& 4 & 4  \\
				3&4 & 5 
			\end{bmatrix}.
		\end{equation}
		We then obtain $ \mathbf{a}_B =[1,1,1,1,1,0,1,0,1]^T$ and the corresponding $ \mathbf{R} \in \mathbb{R}^{9\times 5}=\left[
		1,0,0,0,0\right.$; $0,1,0,0,0 $;
		$  0,0,1,0,0$; $0,1,0,0,0 $;
		$0,0,0,1,0$; $0,0,0,0,0$;
		$  0,0,1,0,0$;$0,0,0,0,0 $;
		$\left.0,0,0,0,1
		\right]$ based on \eqref{eq:R_m}. 
		%
		Since $ \widetilde{\mathbf{b}}=\operatorname{vec}(\mathbf{B}) $ and $ \mathbf{b} = \operatorname{vec}_i(\mathbf{B})= [[\mathbf{B}]_{1,1},[\mathbf{B}]_{1,2},[\mathbf{B}]_{1,3},[\mathbf{B}]_{2,2},[\mathbf{B}]_{3,3}] $, $ \widetilde{\mathbf{b}} =\mathbf{R}\mathbf{b} $ is obtained.
		\section{Proof for proposition \ref{pro:fully-con}}\label{app:proof-fully}
		When $Q = N - 1$, i.e., for a fully-connected RIS, problem~\eqref{eq:channelgain-formula} reduces to:
		\begin{subequations}    
			\label{eq:fully}
			\begin{align}
				\max_{\bm \Theta} \quad &\|\mathbf{S}_{M}\mathbf{V}_{M}^H\mathbf \Theta \mathbf{P}_{M}\bm \Sigma_M\|_F^2\\
				\operatorname{s.t.} \quad &\bm\Theta\bm\Theta^H=\mathbf I_N, \bm\Theta=\bm\Theta^T.
			\end{align}
		\end{subequations}
		
		The theoretical upper bound of problem \eqref{eq:fully} is attainable if and only if the following equation set has solutions
		\begin{subequations}\label{sys of eq}
			\begin{align}
				\label{sys1}  \mathbf{V}_{M}^H\mathbf \Theta \mathbf{P}_{M}&=\bm\Phi,\\
				\label{sys2} \mathbf \Theta\mathbf \Theta^H&=\mathbf I_N,\\
				\mathbf \Theta&=\mathbf \Theta^T.  
			\end{align}
		\end{subequations}
		Based on \eqref{eq:theta-symplified} for \eqref{sys1}-\eqref{sys2},  \eqref{sys of eq} is further simplified as
		\begin{equation}\label{simplified sys}
			\bm \Theta=  \mathbf{V}_{M} \mathbf{P}_{M}^H + \mathbf{V}_{N-M} \mathbf{X}_0\mathbf{P}_{N-M}^H,
			\mathbf \Theta=\mathbf \Theta^T. 
		\end{equation}
		When considering the special case $ M=1 $, \eqref{simplified sys} has solutions as shown in \cite{nerini2023closed, takagi2023}. Next, we prove  by contradiction that when  $ M>1 $,
		\eqref{simplified sys} has no solutions.
		
		Assume that there exists at least one  solution for \eqref{simplified sys}, then
		\begin{equation}\label{eq:should-sym}
			\begin{split}
				&\mathbf{V}_{M} \mathbf{P}_{M}^H + \mathbf{V}_{N-M} \mathbf{X}_0\mathbf{P}_{N-M}^H\\
				=&\mathbf{P}_{M}^*\mathbf{V}_{M}^T + \mathbf{P}_{N-M}^*\mathbf{X}_0^H\mathbf{V}_{N-M}^T .
			\end{split}
		\end{equation}
		By left multiplying $\mathbf{V}_{M}^H$ and right multiplying $\mathbf{V}_{M}^*$ to \eqref{eq:should-sym}, we have $ \mathbf{P}_{M}^H \mathbf{V}_{M}^* = \mathbf{V}_{M}^H \mathbf{P}_{M}^* $.
		Let $\bm \Lambda \triangleq \mathbf{P}_{M}^H \mathbf{V}_{M}^* \in \mathbb{C}^{M \times M}$, we have $\bm \Lambda=\bm \Lambda^T$. When $M=1$, $\bm \Lambda \in \mathbb{C}$ is a constant scalar, so that $\bm \Lambda=\bm \Lambda^T$ holds obviously. However, when $M>1$, $\bm \Lambda$ is not symmetric with probability $1$ because $ \mathbf{V}_{M} $ and $ \mathbf{P}_{M} $ are unitary matrices from the SVD of $ \mathbf{H}^H $ and $ \mathbf{E} $, respectively. This contradicts our initial assumption.
		Therefore, when $ M>1 $, no solution for \eqref{sys of eq} exists. Hence, when $ M>1 $, fully connected RIS cannot achieve the theoretical upper bound performance, completing the proof for the proposition.
		
	\end{appendices}
	\bibliographystyle{IEEEtran}  
	\bibliography{reference}

\end{document}